\documentclass[a4paper,11pt]{article}

\usepackage{youngtab}
\usepackage{graphicx,array}
\usepackage{color}
\usepackage{latexsym}
\usepackage{amsthm}
\usepackage{amsmath}
\usepackage{amssymb}
\usepackage{empheq}
\usepackage{mathtools}
\usepackage{scalerel,amssymb}

\usepackage{dsfont}
\usepackage{epsfig}
\usepackage{slashed}
\usepackage{bbold}
\usepackage{psfrag}
\usepackage[svgnames]{xcolor}
\PassOptionsToPackage{caption=false}{subfig}
\usepackage{subcaption}
\usepackage{xfrac}
\usepackage{multirow}
\usepackage{booktabs}
\usepackage{cite}
\usepackage[normalem]{ulem}
\usepackage{jheppub} % for details on the use of the package, please see the JHEP-author-manual
\usepackage{colortbl}
\usepackage{array,booktabs,arydshln,xcolor}
\newcommand\VRule[1][\arrayrulewidth]{\vrule width #1}

\newcommand{\be}{\begin{equation}}
\newcommand{\ee}{\end{equation}}
\newcommand{\bea}{\begin{eqnarray}}
\newcommand{\eea}{\end{eqnarray}}
\newcommand{\bei}{\begin{itemize}}
\newcommand{\eei}{\end{itemize}}

\newcommand{\nn}{\nonumber}

\title{Perturbative unitarity constraints on generic Yukawa interactions}

\author[a]{Lukas Allwicher}
\author[b,c]{Pere Arnan}
\author[b,c]{Daniele Barducci}
\author[b,c]{Marco Nardecchia}

\affiliation[a]{Physik-Institut, Universit\"at Z\"urich, CH-8057 Z\"urich, Switzerland
}
\affiliation[b]{Universit\`a degli Studi di Roma la Sapienza, Piazzale Aldo Moro 5, 00185, Roma, Italy}
\affiliation[c]{INFN Section of Roma 1, Piazzale Aldo Moro 5, 00185, Roma, Italy}

\emailAdd{lukall@physik.uzh.ch}
\emailAdd{pere.arnan@roma1.infn.it}
\emailAdd{daniele.barducci@roma1.infn.it}
\emailAdd{marco.nardecchia@roma1.infn.it}

\abstract{We study perturbative unitarity constraints on generic Yukawa interactions where the involved fields have arbitrary quantum numbers under an $\prod_i SU(N_i) \otimes U(1)$ group. We derive compact expressions for the bounds on the Yukawa couplings for the cases where the fields transform under the trivial, fundamental or adjoint representation of the various $SU(N)$ factors. We apply our results to specific models formulated to explain the anomalous measurements of $(g-2)_\mu$ and of the charged- and neutral-current decays of the $B$ mesons. We show that, while these models can generally still explain the observed experimental values, the required Yukawa couplings are pushed at the edge of the perturbative regime.
}

\begin{document} 
\maketitle

%%%%%%%%%%%%%%%%%
%%%	INTRO		   %%%		
%%%%%%%%%%%%%%%%%

\section{Introduction}

Yukawa interactions are, together with gauge and scalar self-interactions, the building blocks of renormalizable theories. In the Standard Model (SM) they are a crucial ingredient as they describe the interactions between quarks and leptons and the Higgs boson, with the latter being ultimately responsible for the generation of fermion masses after electroweak symmetry breaking (EWSB).

They are also ubiquitous in new physics (NP) theories that try to address the shortcomings of the SM. For example, they appear in theories that generate neutrino masses both at tree-level, as for the case of the well know seesaw mechanism~\cite{Minkowski:1977sc,Mohapatra:1979ia,Yanagida:1979as,Gell-Mann:1979vob,Schechter:1980gr,Schechter:1981cv,
Lazarides:1980nt,Mohapatra:1980yp,
Foot:1988aq}, as well as at higher orders~\cite{Zee:1980ai,Zee:1985id,Babu:1988ki,FileviezPerez:2009ud} and in models where a fermion Dark Matter (DM) candidate is connected to the SM through a scalar portal, see {\emph{e.g.}}~\cite{Abercrombie:2015wmb} for a review. Interestingly, the existence of new scalar bosons  or some beyond the SM (BSM) fermions that possess Yukawa interactions with the SM could solve some anomalies reported in the recent years in low energy data.
This is, for example, the case of the measurement of the  anomalous magnetic moment of the muon $(g-2)_\mu$, for which the recent measurement by the E989 experiment at Fermilab~\cite{Muong-2:2021ojo}, which is in agreement with the previous BNL E821 result~\cite{Muong-2:2006rrc}, implies a $\sim 4.2\sigma$ discrepancy with respect to the SM prediction~\cite{Aoyama:2020ynm}, although a recent lattice calculation seems in agreement with it~\cite{Borsanyi:2020mff}.  It is also the case of other long-standing anomalies in semileptonic decays of $B$ mesons both in charged-~\cite{BaBar:2012obs,BaBar:2013mob,Belle:2015qfa,Belle:2016ure,Belle:2016dyj,LHCb:2015gmp,LHCb:2017smo,LHCb:2017rln,Belle:2019gij} and neutral-current~\cite{LHCb:2014vgu,LHCb:2017avl,LHCb:2021trn} decays, usually
dubbed $R_{D^{(*)}}$ and $R_{K^{(*)}}$,
that can be accounted for  by various models involving additional Yukawa sectors.
 Some or all of these anomalies can be solved by postulating the existence of leptoquarks (LQ)~\cite{Gripaios:2014tna,Becirevic:2015asa,Varzielas:2015iva,Alonso:2015sja,Calibbi:2015kma,Belanger:2015nma,Barbieri:2015yvd,Becirevic:2016oho,Becirevic:2016yqi,Sahoo:2016pet,Hiller:2016kry,Cox:2016epl,Crivellin:2017zlb,Cai:2017wry,Dorsner:2017ufx,Buttazzo:2017ixm,Assad:2017iib,DiLuzio:2017vat,Calibbi:2017qbu,Bordone:2017bld,Barbieri:2017tuq,Blanke:2018sro,Marzocca:2018wcf,Bordone:2018nbg,Becirevic:2018afm,Kumar:2018kmr,Crivellin:2018yvo,deMedeirosVarzielas:2018bcy,Azatov:2018kzb,DiLuzio:2018zxy,Faber:2018qon,Heeck:2018ntp,Angelescu:2018tyl,Arnan:2019olv,Gherardi:2019zil,Cornella:2019hct,Crivellin:2019dwb,Fuentes-Martin:2020bnh,Saad:2020ihm,Crivellin:2020ukd,Gherardi:2020qhc,Bordone:2020lnb,DaRold:2020bib,Angelescu:2021lln,Greljo:2021xmg,Marzocca:2021azj,Marzocca:2021miv,Greljo:2021npi}, {\emph{i.e.}} new colored  states which connect quark and leptons, or by extending the SM with new heavier scalars and vector-like fermions \cite{Gripaios:2015gra,Arnan:2016cpy,Kawamura:2017ecz,Cline:2017qqu,Barman:2018jhz,Grinstein:2018fgb,Li:2018rax,Li:2019xmi,Cerdeno:2019vpd,Arnan:2019uhr,Huang:2020ris,Arcadi:2021glq,Becker:2021sfd,Arcadi:2021cwg}.
Obviously, new Yukawa interactions are constrained by a large variety of experimental searches, ranging from direct production of new on-shell degrees of freedom at high energy colliders to low energy precision measurements. The results of these analyses are generally expressed as limits on combinations of couplings and masses, and the resulting bounds  strongly depend on the specific structure of the NP realization and on the experimental settings.

There exists, however, and old tool of theoretical physics, namely perturbative unitarity (PU), that can be used to set an upper limit on the magnitude of the couplings, above which the perturbative expansion is expected to break down. Most famously this tool, that we review in Sec.~\ref{sec:pu}, has been applied to set an upper bound on the Higgs boson mass~\cite{Lee:1977yc,Lee:1977eg,Marciano:1989ns,Horejsi:2005da} 
and on the masses of quarks and leptons participating in weak interactions
\cite{Chanowitz:1978mv,Chanowitz:1978uj}~\footnote{See also~\cite{Dicus:2004rg,Dicus:2005ku} for related works.} if weak  interactions were to remain weak at all energies. 
It has then been widely used in the literature to assess the range of validity of both renormalizable
 and effective operators~\cite{Griest:1989wd,Hally:2012pu,Kahlhoefer:2015bea,Chang:2019vez,Abu-Ajamieh:2020yqi,DiLuzio:2017tfn,DiLuzio:2016sur,Capdevilla:2021rwo,Allwicher:2021jkr,DiLuzio:2017chi,Corbett:2014ora,Corbett:2017qgl,Almeida:2020ylr,Brivio:2021fog}.  In this work we consider the problem in more generality and we answer the following question:
 \vspace{-0.085cm}
\begin{quote}
 {\emph{Given a Yukawa interaction between a scalar and two fermions with generic quantum number under a group ${\cal G}=\prod_i SU(N_i) \otimes U(1)$, what is the maximum allowed value for the coupling with the requirement of PU? }}
\end{quote}
To answer this question we  consider the most general form of Yukawa-type interactions and all possible $2\to 2$ tree-level scatterings in the high-energy limit. We obtain compact expressions for the upper limit on the value of the Yukawa coupling up to which perturbation theory could be trusted, and highlight their dependence on the various fields' quantum numbers under ${\cal G}$.

More specifically, we start by computing all the necessary ingredients for building the partial wave scattering matrix, namely the Lorentz parts of the scattering amplitudes and the group structure factors entering the amplitudes themselves, in a set of phenomenologically relevant toy models where the various fields are only charged under a {\emph{single}} $SU(N)$ factor.
We firstly show how the $SU(N)$ group structure of the interaction can lead to en 
enhancement of the scattering amplitudes  and thus to a tightening of the partial wave unitarity bounds, while the role of the $U(1)$ charge is to enforce a selection rule that makes some amplitudes vanish. We then use these toy models as building blocks for more complicated theories, where the various fields are charged under {\emph{multiple}} $SU(N_i)$ factors, giving as a working example the case of the SM quark Yukawa sector.
We then apply our results to different NP models which solve, by introducing a new Yukawa sector, the aforementioned anomalies in $(g-2)_\mu$ and/or in the semileptonic decay of $B$ meson, showing that while the proposed theories can generally still provide an explanation to these measurements, their model parameters are stretched close to the limit of our-tree level unitarity bound criteria.

Altogether the results presented in this work are of practical use and can be used to analyze scenarios beyond the examples presented in the text.  While we restrict only to a limited number of irreducible $SU(N)$ representations under which the various fields can transform (singlet, fundamental and adjoint), we believe that our computations furnish the necessary ingredients to study a large set of NP theories with additional Yukawa interactions.

The paper in organized as follows. In Sec.~\ref{sec:pu} we review the tool of PU and clarify the physical interpretation of the inferred bounds, while in Sec.~\ref{sec:gen} we discuss the general properties of Yukawa interactions relevant for the study of PU. In Sec.~\ref{sec:toy} we introduce the toy models and discuss how they can be used to construct the most general partial wave scattering matrix. Then in Sec.~\ref{sec:gauge-dirac} we study the partial wave unitarity bounds for the first type of toy models which have a {\emph{Dirac type}} structure for the Yukawa interaction. In Sec.~\ref{sec:sm_yuk} we apply
our formalism to the case of the SM quark Yukawa sector, also highlighting the role that multiplicity due to a flavor structure can have in the determination of the bound. In Sec.~\ref{sec:gauge-majo} we then discuss a second class of toy models, which present {\emph{Majorana type}} Yukawa interactions.
Then in Sec.~\ref{sec:pheno} we show some phenomenological applications, finally  concluding in Sec.~\ref{sec:conc}.  We also add few relevant appendices. In App.~\ref{app:conventions} we list our conventions for the calculation of the partial wave matrix, while in App.~\ref{app:dirac} we report the results for other {\emph{Dirac type}} theories not included for brevity in the main text.

%%%%%%%%%%%%%%%%%
%%%	PU TOOL		   %%%		
%%%%%%%%%%%%%%%%%

\section{General aspects}

\subsection{The tool of perturbative unitarity}\label{sec:pu}

Perturbation theory is a powerful tool to provide approximate solutions to physical problems. Within this approach the relevant result is expressed in terms of a power series of some small parameter $\epsilon$. In general, however, it is not easy to determine if some specific numerical value of $\epsilon$ allows for a good approximate solution of the problem under consideration. The aspect we want to face in this Section is to identify a reasonable criterium to state whether a parameter is too large to be treated in perturbation theory.

A level zero criterium for considering a parameter as perturbative is to ask  the expansion parameter entering the $\beta$ functions $\epsilon=\frac{g^2}{(4\pi)^2}$, to be small. The requirement $\epsilon<1$ implies $g<4\pi$, which is the maximum value frequently adopted in the literature.
This condition can however be improved by analyzing the problem in more detail. Consider an abelian gauge theory with coupling $g$ and $N_f$ copies of matter fields charged under the local symmetry. The  scaling of the contribution to the one-loop two point function for the gauge field is
\begin{center}
\includegraphics[width=0.55\textwidth]{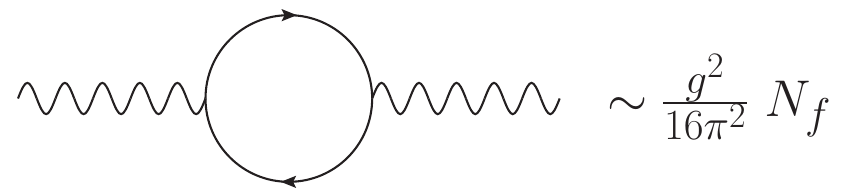}.
\end{center}
The requirement of small expansion parameter then suggest a more refined version of the naive criterium, since a small value of $g$ is in fact not enough if $N_f$ becomes very large. One can then require $g < \frac{4 \pi}{\sqrt{N_f}}$, where it is clear that the multiplicity of the matter fields has to be taken into account for a more refined version of the perturbative criterium. One can however do even better than the former proposal. It is possible to use results that hold beyond perturbation theory to motivate a more stringent criterium based on partial wave PU.

The key point of our analysis are the so-called partial waves, {\emph{i.e.}} the scattering amplitudes with fixed total angular momentum $J$. In the case of $2\to 2$ scatterings in the high-energy massless limit they are defined as~\cite{Jacob:1959at} 
\be\label{eq:partial_waves}
a_{fi}^J = \frac{1}{32\pi}\int_{-1}^1 {\rm d}\cos\theta d_{\mu_i\mu_f}^J(\theta) {\cal T}_{fi}({\sqrt s, \cos\theta}) \ .
\ee
Here $\theta$ is the polar scattering angle in the center of mass frame and $\sqrt s$ the center of mass energy, $d_{\mu_i\mu_f}(\theta)$ are the small Wigner $d-$functions where $\mu_i=\lambda_{i_1}-\lambda_{i_2}$ and $\mu_f=\lambda_{f_1}-\lambda_{f_2}$ are defined in terms of the helicities of the initial and final states, and $(2\pi)^4\delta^{(4)}(P_i -P_f)i {\cal T}_{fi}(\sqrt s, \cos\theta)=\langle f| S-1| i \rangle$, with $S$ the $S$-matrix, that defines the scattering amplitude. We report in App.~\ref{app:wig} the definition and the explicit expressions of the small Wigner $d-$functions used throughout our analysis. The unitarity condition on the $S-$matrix, $S^\dag S = 1$, implies 
\be
\frac{1}{2i} (a_{fi}^J-a_{if}^{J*})=\sum_h a_{hf}^{J*}a_{hi}^J \ ,
\ee
where the sum runs over all the intermediate states $h$. By focusing on elastic channels $i=f$ and restricting the sum over $h$ only to $2$-particle states one obtains the condition
\be
{\rm Im}[a_{ii}^J] \ge |a_{ii}^J|^2 \ .
\ee
This last equation defines a circle in the complex plane, the Argand circle, inside which the amplitude must lie {\emph{at all orders in perturbation theory}}
\be
{\rm Re}^2[a_{ii}^J] +\left({\rm Im}[a_{ii}^J]-\frac{1}{2}\right)^2 \le \frac{1}{4} \ .
\ee
Since for $2\to 2$ high-energy scatterings the tree-level elastic amplitudes are real\footnote{This is a consequence of the optical theorem: intermediate states cannot go on-shell at the tree level if the scattering energy is much larger than their masses.}, this suggests the following unitarity bound
\be\label{eq:pu-limit}
{\rm Re}(a_{ii}^{J,{\rm tree}}) \le \frac{1}{2} \ .
\ee
While the factor $1/2$ is somewhat arbitrary, it gives a reasonable indication of the range of validity of the perturbative expansion, since a tree-level value which saturates Eq.~\eqref{eq:pu-limit} needs at least a higher-order correction of $\sim 40\%$ in order to re-enter the Argand circle, thus signalling the breakdown of the perturbative expansion itself. In order to extract the best PU bound, one then needs to identify the optimal elastic channel and this corresponds to 
diagonalizing the partial wave scattering matrix $a^J_{fi}$. The most stringent limit will be then set by the largest, in absolute value, eigenvalue.

%%%%%%%%%%%%%%%%%%
%%%	GENERAL ASPECTS		   %%%		
%%%%%%%%%%%%%%%%%%

\subsection{The structure of the Yukawa interaction}\label{sec:gen}
In this section we specify the class of models we are interested in and the assumptions we make in our analysis. 

First of all we assume that mass terms are negligible and all the computation are performed in the high energy regime. 
In this limit we can consider massless fields as the most natural degrees of freedom. Regardless of the symmetry structure of the interactions, we can always (re-)write a generic Yukawa interaction between a set of $N_\phi$ real scalar fields $\phi_\alpha$ and of $N_\psi$ Weyl fermion fields $\psi_{L}^i$ in the following way:
\be\label{eq:gen_lag}
- {\cal L} = \frac{1}{2} {\cal Y}_{\alpha i j} 
\phi_\alpha \bar \psi_L^i   \psi_L^{c,j} + h.c. \ ,
\ee
where $\psi^c_{L,i}=C \bar \psi^T_{L,i}$,  $C=i \gamma^2 \gamma^0$ is the charge conjugation matrix, ${\cal Y}_{\alpha ij}={\cal Y}_{\alpha ji}$ and the index $\alpha$ runs from 1 to $N_\phi$, while $i,j = 1,\ldots N_\psi$. 
Notice that this form is the most general one. For example, a complex scalar field can be always expressed in terms of two real fields with specific restrictions on the phases of the coupling  ${\cal Y}_{\alpha ij}$, in a similar way symmetry properties of the Yukawa interactions are manifest through the presence of null elements or by specific relations among them. We are going to clarify these aspects in what follows with explicit toy models.

Our task is now to compute the partial wave matrices $a_{fi}^J$ of Eq.~\eqref{eq:partial_waves} and extract their eigenvalues. We start discussing the helicity and Lorentz structure of the scattering amplitudes. To this end we firstly need to compute the ${\cal T}_{fi}(\sqrt s,\cos\theta)$ amplitudes between the initial and final states. It is useful to write the scattering matrix ${\cal T}$ in the following form
\vskip 10pt
\be\label{eq:full_matrix}
    {\cal T}_{fi} =
     \left(
    \begin{array}{c c !{\VRule[1pt]}c c|c|c|c c}
           &     &   \multicolumn{2}{c|}{\mu_i=0} &  \mu_i=0 &  \mu_i=+1 &  \mu_i=+1/2 &  \mu_i=-1/2 \\
         &  &  ++ 				& -- & 00 & +- & +0 & -0 \\
\specialrule{1pt}{0pt}{0pt}
\multirow{2}{*}{$\mu_f=0$} & ++   & {\cal T}^{++++}	 & {\bf {\cal T}}^{++--}  &  &  &  &  \\
 &--   & {\cal T}^{--++} 	 & {\bf {\cal T}}^{----}  &  &  &  &  \\
\hline
\mu_f=0 &00   &  	 &  &  & {\bf {\cal T}}^{00+-} &  &  \\
\hline
\mu_f=+1 &+-   &  	&   &{\bf {\cal T}}^{+-00}  & {\bf {\cal T}}^{+-+-} &   & \\
\hline
\mu_f=+1/2 &+0  &  	&   &  &   & {\bf {\cal T}}^{+0+0} & \\
\mu_f=-1/2  &+0  &  	&   &  &   & & {\bf {\cal T}}^{-0-0} \\
    \end{array}
    \right) \ ,
\ee
\vskip 10pt
\noindent where we have indicated the total helicities of the initial and final states, $\mu_{i,f}$, as well as the helicities of the single particles involved in the scattering. Each of the amplitudes entering the various blocks of the ${\cal T}_{fi}$ matrix are themselves matrices, whose dimensionalities depend on the number of fermions and scalars of a given theory. Notice that the matrix of 
Eq.~\eqref{eq:full_matrix} has many zero entries, which correspond to the empty blocks. In particular the ${\cal T}^{\pm0\pm\pm}$, ${\cal T}^{\pm0+-}$  and ${\cal T}^{\pm000}$ 
amplitudes vanish because of total angular momentum conservation, while ${\cal T}^{0000}$ is present if one adds also a potential for the scalar fields. Finally, the other null amplitudes are strictly zero only in the massless limit we are considering. Since we are working in the high-energy massless limit it is useful to compute the non vanishing amplitudes by working with helicity eigenstates, following the conventions of Jacob and Wick~\cite{Jacob:1959at}. We refer the reader to App.~\ref{app:hel} for the details on the choice of the spinor helicity basis.  The remaining non vanishing amplitudes can be computed explicitly starting from the interaction in Eq.~\eqref{eq:gen_lag}. 
All the non vanishing amplitudes are reported in Eq.~\eqref{eq:amp}

The usefulness of Eq.~\eqref{eq:full_matrix} is that, when projecting the amplitudes onto the $J-$th partial wave via Eq.~\eqref{eq:partial_waves}, only a subset of the non-zero blocks survives. In particular the channels with $\mu_i=\mu_f=0$ have a non-zero projection only on $J=0$, since the relevant amplitudes do not depend on $\theta$, see Eq.~\eqref{eq:amp}. 
Again because of total angular momentum conservation channels with $\mu_{i,f}=\pm \frac{1}{2}$ project only on half-integers values of $J$. Hence, for integer $J>0$ only the channels ${\cal T}^{+-+-}$ and ${\cal T}^{00+-}$ contribute and the scattering matrix is effectively separated in three different blocks, allowing us to consider each of them independently when studying partial waves with different values of $J$. 
Interestingly, we will show that the stronger bound might arise from a partial wave different from $J=0$. In particular in Sec.~\ref{sec:gauge-dirac} we will see that the tighter limit can come from the analysis of $J=\frac{1}{2}$ or $J=1$, while higher partial waves give a weaker bound.

\begin{table}[t]
\begin{center}
\begin{tabular}[t]{ c | c  | c  | c    }
 \multicolumn{4}{c}{{\bf Dirac type}}\\
Model & $\chi_L$ & $\eta_R$ & $S$ \\
\hline
1& ${\tiny{\yng(1)}}_q$ & ${\tiny{\yng(1)}}_{q^\prime}$ & $\mathbf{1}_{q-q^\prime}$  \\
2 & ${\tiny{\yng(1)}}_q$ & $\mathbf{1}_{q^\prime}$ & ${\tiny{\yng(1)}}_{q-q^\prime}$    \\
3& ${\tiny{\yng(1)}}_q$ & ${\tiny{\yng(1)}}_{q^\prime}$ & Adj$_{q-q^\prime}$    \\
4& ${\tiny{\yng(1)}}_q$ & Adj$_{q^\prime}$ & ${\tiny{\yng(1)}}_{q-q^\prime}$    \\
$5^\star$ & ${\tiny{\yng(1)}}_q$ & $\overline {\tiny{\yng(1)}}_{q^\prime}$ & $\overline {\tiny{\yng(1)}}_{q-q^\prime}$
\end{tabular}  
\hspace{2cm}
\begin{tabular}[t]{ c | c    | c }
 \multicolumn{3}{c}{{\bf Majorana type}}\\
Model & $\chi_L$ &  $S$  \\
\hline
1& ${\mathbf 1}_q$  & ${\mathbf 1}_{2q}$   \\
$2^\star$& ${\tiny{\yng(1)}}_q$  & Adj$_{2q}$ \\
\end{tabular}  
\end{center}
\caption{{\emph{Dirac type}} and {\emph{Majorana  type}} theories described by the Lagrangians of Eq.~\eqref{eq:model_I} and Eq.~\eqref{eq:model_II}, respectively. If $S$ is either a singlet or transforms in the adjoint representation of $SU(N)$ and also has a vanishing $U(1)$ charge, then it is a real scalar field. Model 5 of the first class is only present in the case of $SU(3)$, while model 2 of the second class only in the case of $SU(2)$.}
\label{tab:models}
\end{table}

\section{Toy models and scattering amplitudes}\label{sec:toy}

The Lagrangian of Eq.~\eqref{eq:gen_lag} describes the Yukawa interaction between any set of scalar and fermion fields, where the entries of the Yukawa matrix ${\cal Y}_{\alpha i j}$ are at this level completely generic. By assigning definite quantum numbers under a group ${\cal G}=\prod_i SU(N_i)\otimes U(1)$ to the scalar and fermion fields involved in the interaction, the Yukawa matrix, and consequently the scattering matrix, acquires a well definite structure. In this Section we consider two types of toy models which we use as building blocks for the study of more general Yukawa theories. More specifically we consider two theories, described by the following Lagrangians:
\be\label{eq:model_I}
-{\cal L}_{\rm Dirac} = y S \bar \chi \eta + h.c.  \ ,
\ee
and
\be\label{eq:model_II}
-{\cal L}_{\rm Majorana} = \frac{1}{2} y S \bar \chi \chi^c + h.c.  \ ,
\ee
where $\chi$ and $\eta$ are left-handed and right-handed fermion fields respectively, and $S$ a scalar field that can be either complex or real. We dub these theories as {\emph{Dirac type}} and {\emph{Majorana type}} respectively.
To begin our study, we start by assuming that all fields are charged under a {\emph{single}} $SU(N)$ factor and show later in Sec.~\ref{sec:sm_yuk} how the case of multiple $SU(N_i)$ charges can be dealt with.  For both theories the parameter $y$ can be chosen to be real without loss of generality by a proper field redefinition. For concreteness we restrict our study to the case where all the fields transform in the trivial, fundamental or adjoint representation of $SU(N)$ and we allow them to have arbitrary $U(1)$ charges. For simplicity we also consider theories where at most one field transforms in the adjoint $SU(N)$ representation. Under these assumptions the various models that can be written are reported in Tab.~\ref{tab:models}. Note that in the {\emph{Dirac type}} class, model number 5 can only be written in the case of $SU(3)$ while in the {\emph{Majorana type}} class, model number 2 can only be written in the case of $SU(2)$. A comment regarding other possible models in the {\emph{Majorana type}} class is in order:
\begin{itemize}
\item For the case of $SU(2)$ one can write a gauge invariant interaction with $\chi \sim {\tiny{\yng(1)}}_q$ and $S \sim \mathbf{1}_{2q}$ which however  identically vanishes since $\bar \chi \varepsilon\chi^c=0$, where $\varepsilon$ is the $SU(2)$ totally antisymmetric tensor. One can restore this interaction by charging the field $\chi$ under a second $SU(N)$ factor, or by considering different flavors for $\chi$, since in this case one can antisymmetrize in the additional gauge and/or flavor index.
\item For the same reason, in the case of $SU(3)$ also the model where  $\chi\sim{\tiny{\yng(1)}}_q$ and $S\sim{\tiny{\overline{\yng(1)}}}_{2q}$ vanishes if the states are not charged under another $SU(N)$ group or flavor multiplicity is not added.
\end{itemize}
We will comment on these possibilities when considering in more detail the {\emph{Majorana type}} class of models, presenting in Sec.~\ref{sec:qq} a phenomenologically relevant example in which the involved fields are charged under more than one $SU(N_i)$ factor.

For any given toy model, the task is to to build the $a_{fi}^J$ partial wave matrices and compute their largest eigenvalues. This can be done mechanically by {\emph{brute force}} by building the Yukawa matrix ${\cal Y}_{\alpha i j}$ entering Eq.~\eqref{eq:gen_lag}, computing all the amplitudes\footnote{Here one has to consider the relevant factors of $1/\sqrt{2}$ for identical particles, which can occur only for two-fermion states when $\mu_i$ and/or $\mu_f=0$ and for two-scalar states.} and building the $a_{fi}^J$ matrices explicitly through Eq.~\eqref{eq:full_matrix}.
Although straightforward, this process turns out to be highly inefficient, due to the rapid increase in the $a_{fi}^J$ matrix dimension when considering $SU(N)$ factors with large $N$. As an example for the third theory of the {\emph{Dirac type}} with a complex scalar field, the transition matrix has dimension 406 for $N=3$ and  2346 for $N=5$. When considering the possibility of fields charged under more than one $SU(N_i)$ factor the dimensionality of the transition matrix dramatically increases, making also the numerical calculation inefficient.

\begin{table}[t!]
\begin{center}
\scalebox{0.8}{
\begin{tabular}{| c|  c ||  c  | c  |  c |  c  | c | c| }
 \multicolumn{8}{c}{{\bf Dirac type models $-{\cal L}=y\bar\chi \eta S + h.c.$}} \\
\hline
  & & \multicolumn{3}{c|}{Real $S$} &   \multicolumn{3}{c|}{Complex $S$}\\
\hline
$\lambda_{f_1}\,\lambda_{f_2}\,\lambda_{i_1}\,\lambda_{i_2}$		& States	& ${\cal T}_s$ & ${\cal T}_t$ & ${\cal T}_u$ & ${\cal T}_s$ & ${\cal T}_t$ & ${\cal T}_u$\\
\hline
$++++$				& $\bar\chi\eta\to \bar\chi\eta$ & $-1$ & &  & $-1$& & \\
$----$				& $\chi\bar\eta\to\chi\bar\eta$ & $-1$ & &  & $-1$& & \\
\hline						
\multirow{3}{*}{$--++$}	& $\bar\chi\eta\to \chi\bar\eta$ & $+1$ & & $+1$ & & & \\
					& $\bar\chi\bar\chi\to \bar\eta\bar\eta$ &  &+1& $+1$ & & & \\
					& $ \eta\eta \to \chi\chi$&  &+1& $+1$ & & & \\					
					\hline						
\multirow{3}{*}{$++--$}	& $\chi\bar\eta\to \bar\chi\eta$ & $+1$ & & $+1$ & & & \\
					& $\chi\chi\to \eta\eta$ &  &+1& $+1$ & & & \\
					& $\bar\eta\bar\eta \to \bar\chi\bar\chi$ &  &+1& $+1$ & & & \\					
\hline		%	
\hline						
\multirow{2}{*}{$+0+0$}	& $\bar\chi S \to \bar\chi S$ & $-\cos{\frac{\theta}{2}}$ & & $-\frac{1}{\cos{\frac{\theta}{2}}}$ & $-\cos{\frac{\theta}{2}}$ & & \\
					& $\eta S \to \eta S$ & $-\cos{\frac{\theta}{2}}$ & & $-\frac{1}{\cos{\frac{\theta}{2}}}$ & $-\cos{\frac{\theta}{2}}$ & & \\		
\hline						
\multirow{2}{*}{$-0-0$}	& $\chi S \to \chi S$ & $-\cos{\frac{\theta}{2}}$ & & $-\frac{1}{\cos{\frac{\theta}{2}}}$ &  & &$-\frac{1}{\cos{\frac{\theta}{2}}}$ \\
					& $\bar\eta S \to\bar \eta S$ & $-\cos{\frac{\theta}{2}}$ & & $-\frac{1}{\cos{\frac{\theta}{2}}}$ &  & & $-\frac{1}{\cos{\frac{\theta}{2}}}$\\			
\hline						
\multirow{2}{*}{$+0^*+0^*$}	& $\bar\chi S^* \to \bar\chi S^*$ &  & &  & & &$-\frac{1}{\cos{\frac{\theta}{2}}}$ \\
					& $\eta S^* \to \eta S^*$ &  & &  &  & &$-\frac{1}{\cos{\frac{\theta}{2}}}$ \\	
\hline						
\multirow{2}{*}{$-0^*-0^*$}	& $\chi S^* \to \chi S^*$ &  & &  &$-\cos{\frac{\theta}{2}}$  & & \\
					& $\bar\eta S^* \to\bar \eta S^*$ &  & &  &$-\cos{\frac{\theta}{2}}$  & & \\				
\hline
\hline						
\multirow{3}{*}{$+-+-$}	& $\bar\chi\chi \to \eta\bar\eta$ & & & $-1$ &  & & $-1$ \\
					& $\bar\chi\bar\eta \to \bar\chi\bar\eta$ & & & $-1$ &  & & $-1$ \\	
					& $\eta\chi\to\eta\chi$ & & & $-1$ &  & & $-1$ \\						
\hline	
\multirow{3}{*}{$00+-$}	& $\bar\chi\chi \to SS$ & & $\frac{1}{\tan{\frac{\theta}{2}}}$ & $-\tan{\frac{\theta}{2}}$ &  & &  \\
					& $\eta\bar\eta \to SS$ & & $\frac{1}{\tan{\frac{\theta}{2}}}$ & $-\tan{\frac{\theta}{2}}$ &  & &  \\	
					\hline	
\multirow{2}{*}{$00^*+-$}	& $\bar\chi\chi \to SS^*$ & &  &  &  & &  $-\tan{\frac{\theta}{2}}$\\
					& $\eta\bar\eta \to SS^*$ & &  & &  & & $-\tan{\frac{\theta}{2}}$  \\	
\hline	
\end{tabular}  
}
\end{center}
\caption{Lorentz part of the amplitudes for the models of the first class divided by $y^2$. In the complex scalar case also the $0^*0^*+-$ amplitudes are zero. In the helicity amplitudes the notation 
$0^*$ indicates the scattering involving the conjugate of the scalar $S$.} 
\label{tab:lorentz_dirac}
\end{table}

\begin{table}[t!]
\begin{center}
\scalebox{0.8}{
\begin{tabular}{| c|  c || c  | c  |  c |  c  | c | c| }
 \multicolumn{8}{c}{{\bf Majorana type models $-{\cal L}=\frac{y}{2}\bar\chi \chi^c S + h.c.$}} \\
\hline
  & & \multicolumn{3}{c|}{Real $S$} &   \multicolumn{3}{c|}{Complex $S$}\\
\hline
$\lambda_{f_1}\,\lambda_{f_2}\,\lambda_{i_1}\,\lambda_{i_2}$		& States	& ${\cal T}_s$ & ${\cal T}_t$ & ${\cal T}_u$ & ${\cal T}_s$ & ${\cal T}_t$ & ${\cal T}_u$\\
\hline
$++++$				& $\bar\chi\bar\chi \to \bar\chi\bar\chi$ & $-1$ & &  & $-1$& & \\
$----$				& $\chi\chi\to\chi\chi$ & $-1$ & &  & $-1$& & \\
\hline						
$--++$	& $\bar\chi\bar\chi\to \chi\chi$ & $+1$ & $+1$ & $+1$ & & & \\
										
\hline						
$++--$	& $\chi\chi\to \bar\chi\bar\chi$ & $+1$  &+1& $+1$ & & & \\

\hline		%	
\hline						
$+0+0$	& $\bar\chi S \to \bar\chi S$ & $-\cos{\frac{\theta}{2}}$ & & $-\frac{1}{\cos{\frac{\theta}{2}}}$ & $-\cos{\frac{\theta}{2}}$ & & \\
		
\hline						
$-0-0$	& $\chi S \to \chi S$ & $-\cos{\frac{\theta}{2}}$ & & $-\frac{1}{\cos{\frac{\theta}{2}}}$ &  & &$-\frac{1}{\cos{\frac{\theta}{2}}}$ \\
			
\hline						
$+0^*+0^*$	& $\bar\chi S^* \to \bar\chi S^*$ &  & &  & & &$-\frac{1}{\cos{\frac{\theta}{2}}}$ \\
			
\hline						
$-0^*-0^*$		& $\chi S^* \to \chi S^*$ &  & &  &$-\cos{\frac{\theta}{2}}$  & & \\
								
\hline		%	
\hline						
$+-+-$	& $\bar\chi\chi \to \bar\chi\chi$ & & & $-1$ &  & & $-1$ \\
	
\hline	
$00+-$	& $\bar\chi\chi \to SS$ & & $\frac{1}{\tan{\frac{\theta}{2}}}$ & $-\tan{\frac{\theta}{2}}$ &  & &  \\

\hline	
$00^*+-$	& $\bar\chi\chi \to SS^*$ & &  &  &  & &  $-\tan{\frac{\theta}{2}}$\\
					
\hline														
\end{tabular}  
}
\end{center}
\caption{Lorentz part of the amplitudes for the models of the second class divided by $y^2$. In the complex scalar case also the $0^*0^*+-$ amplitudes are zero. In the helicity amplitudes the notation 
$0^*$ indicates the scattering involving the conjugate of the scalar $S$.}
\label{tab:lorentz_majorana}
\end{table}

The situation drastically simplifies if one realizes that when considering any $2\to 2$ scattering, each amplitude can be decomposed into a Lorentz part which depends only on the spin and helicity of the involved fields, and a group-theoretical part that depends on their
$SU(N)$ quantum numbers, while the role of the $U(1)$ charge is to enforce a selection rule that will make some amplitudes vanish.  More concretely any $2\to 2$ scattering amplitude among particles $i_{1,2}$ and $f_{1,2}$ with helicities $\lambda_{i_{1,2}}$ and $\lambda_{f_{1,2}}$ can be written schematically as
\be\label{eq:total_T}
{\cal T}_{f_1 f_2 i_1 i_2}^{\lambda_{f_1}\lambda_{f_2}\lambda_{i_1}\lambda_{i_2}} (\sqrt s, \theta)=\bigoplus_{{\bf{r}}} \sum_{m=s,t,u} {\cal T}^{\lambda_{f_1}\lambda_{f_2}\lambda_{i_1}\lambda_{i_2}}_m(\sqrt s,\theta) {\cal F}^{m, {\rm {\bf{r}}}}_{f_1 f_2 i_1 i_2} (N) {\mathbb 1}_{d_{{\bf{r}} }}\ ,
\ee
where $ {\cal T}^{\lambda_{f_1}\lambda_{f_2}\lambda_{i_1}\lambda_{i_2}}_m(\sqrt s,\theta) $ is the Lorentz part of the scattering amplitude and ${\cal F}^{m, {\rm {\bf{r}}}}_{f_1 f_2 i_1 i_2} (N)$ is a function that contains the group part coefficient for the scattering through the Mandelstam $m-$channel  in the $SU(N)$ ${\bf r}$ irreducible representation that can be built from the initial and final state particles, while $d_{\bf{r}}$ stands for the dimensionality of ${\bf r}$.
The direct sum runs over all the irreducible representation through which a scattering can proceed. One of the necessary ingredients are thus the $ {\cal T}^{\lambda_{f_1}\lambda_{f_2}\lambda_{i_1}\lambda_{i_2}}_m(\sqrt s,\theta) $ functions for the two theories of Eq.~\eqref{eq:model_I} and Eq.~\eqref{eq:model_II}, which can be computed from  the Lagrangian of Eq.~\eqref{eq:gen_lag}.  We report them, normalized by the common $y^2$ factor, in Tab.~\ref{tab:lorentz_dirac} and Tab.~\ref{tab:lorentz_majorana} for both the real and complex scalar $S$ case\footnote{As mentioned in Sec.~\ref{sec:gen} we find now convenient, in the case of a complex scalar field, to  directly work with its complex components instead of the  real ones, since as we will see the presence of the $U(1)$ symmetry allows us to simplify the scattering structure.}. Here we clearly see the role played by the $U(1)$ factor. As an example, when $S$ is a complex scalar field the amplitude $\chi \bar{\eta} \to  \bar \chi \eta$ in ${\cal T}^{++--}$ is zero, because this process violates the conservation of the $U(1)$ charge. Analogous selection rules appear in other scattering channels.

In order to fix the idea let us make an explicit example and consider the first theory of the {\emph{Dirac type}} class, where both $\chi$ and $\eta$ transform under the fundamental representation of $SU(N)$ and $S$ is a scalar singlet. The  scattering $\bar\chi \eta \to \bar \chi \eta$ which proceeds through the $++++$ helicity channel has only an $s-$channel contribution, with an amplitude which is proportional to $-y^2$, see Tab.~\ref{tab:lorentz_dirac}. Since ${\tiny{\yng(1)}}\otimes {\overline{{\tiny{\yng(1)}}}}=\mathbf{1}+{\rm Adj}$, this scattering can only proceed through the singlet and the adjoint channels. Thus the scattering amplitude, by applying Eq.~\eqref{eq:total_T},  reads\footnote{In this particular case the amplitude in the adjoint channel vanishes since the amplitude has only an $s-$channel contribution and  $S$ is a $SU(N)$ singlet.}
\be
{\cal T}_{\bar\chi \eta  \bar\chi \eta}^{++++}(\sqrt s,\theta)  = - y^2 
\begin{pmatrix}
{\cal F}^{s,\mathbf{1}}_{\bar\chi \eta  \bar\chi \eta}(N) & \\
& {\cal F}^{s,{\rm Adj}}_{\bar\chi \eta  \bar\chi \eta}(N) {\mathbb 1}_{N^2-1}
\end{pmatrix} \ .
\ee
With the same procedure one can build the amplitudes among irreducible representations for all the possible scatterings of the theory. In each of the separated subsectors that we have identified ($J=0$, half-integer $J$ and integer $J>0$, see Eq.~\eqref{eq:full_matrix}), the matrix can be decomposed into scattering blocks among the various irreducible $SU(N)$ representations. 
Barring the convolution with the Wigner $d-$functions and the integration over the angular variable $\theta$, finding the eigenvalues of the partial wave matrix, and thus extracting the partial wave unitarity bound, is then a trivial task. One only needs to compute the ${\cal F}^{m, {\rm {\bf{r}}}}_{f_1 f_2 i_1 i_2} (N)$ factors.
 The advantage of this procedure with respect to the mechanical {\emph{brute force}}  one previously described is clear: being the group factor proportional to the identity in group space, one needs in practice to consider for each representation only one scattering among the $d_{\bf{r}}$ ones, since all of them will give the same result\footnote{The Mathematica package {\tt SARAH}~\cite{Goodsell:2020rfu} performs in automatic way a similar decomposition for pure scalar theories with an $SU(N)$ group symmetry.}.

\section{Dirac type theories}\label{sec:gauge-dirac}

In the previous Section we have described the general strategy for computing the $a_{fi}^J$ partial wave matrices and presented the Lorentz part of the amplitudes that are needed to compute them. In this Section we compute, for the various models of the {\emph{Dirac type}} class presented in Tab.~\ref{tab:models}, the ${\cal F}^{m, {\rm {\bf{r}}}}_{f_1 f_2 i_1 i_2} (N)$ group factors. For brevity of presentation we report in the main text only the results for the first two type of models, while we defer to App.~\ref{app:dirac} for the remaining ones.

\subsection{First model: $\chi\sim {\tiny{\fbox{$\protect\phantom s$} }}_q$, $\eta\sim {\tiny{\fbox{$\protect\phantom s$} }}_{q^\prime}$ $S\sim \mathbf{1}_{q-q^\prime}$ }\label{sec:model1}

In this model $S$ is an $SU(N)$ singlet, real if $q=q^\prime$, while $\eta$ and $\chi$ transform under the fundamental representation of the $SU(N)$ group.
By choosing as basis\footnote{We adopt $SU(N)$ tensor notation  where lower (upper) indices transform in the fundamental (conjugate) $SU(N)$ representation.  }
\be
\psi_L = (\chi_{a},\eta^{c,a})^T\ , \qquad
\phi=
\begin{cases}
S \qquad\qquad\quad  {\rm Real~scalar}\\
(S,S^*)^T \qquad  {\rm \, Complex~scalar}\\
\end{cases} \ ,
\ee
where $a$ runs from 1 to $N$. The Yukawa matrix of Eq.~\eqref{eq:gen_lag} thus reads
\be
{\cal Y}_\alpha= y
\begin{pmatrix}
& \mathbb{1}_N \\
\mathbb{1}_N& 
\end{pmatrix} \,,
\ee
where $\alpha=1$ in the real scalar case and $\alpha=1,2$ in the complex scalar one. As previously stressed we can separately consider the sectors with $J=0$,  half-integer and integer $J>0$, since no amplitude has a non zero projection on more than one sector. Let us start by considering  $J=0$ with a real scalar $S$.  The two particle states with $++$ helicities decompose as
\be\label{eq:model1_J0_dec}
++
\begin{cases}
 \bar\chi \eta  \sim {\bf 1}+ {\rm Adj}  \\
 \eta \eta \sim {\rm{\bf S}}+{\rm{\bf AS}} \\
 \bar\chi\bar \chi \sim  {\rm{\bf{\overline{S}}}}+{\rm{\bf \overline{AS}}} , 
\end{cases} 
\ee
while the states with $--$ helicities are their conjugates. Here above we indicate with ${\rm{\bf S}}$ and ${\rm{\bf AS}}$ the totally symmetric and antisymmetric irreducible representations that arise from the tensor decomposition ${\tiny{\yng(1)}}\otimes {\tiny{\yng(1)}} = {\rm{\bf S}} \oplus {\rm{\bf AS}}$. Note, however, that the antisymmetric combination of two identical fermions identically vanishes for even $J$, see {\emph{e.g.}}~\cite{Jacob:1959at}. In order to compute the scattering amplitudes we need to explicitly write the two-particle states. We define them as 
\begin{align}\label{eq:states}
& |\psi \bar \psi \rangle_{\bf 1} =  \frac{1}{\sqrt N}\delta_a^b|\psi_a \bar \psi^b\rangle \nn \\
& |\psi \bar \psi \rangle^A_{{\rm Adj}} = \sqrt{2}(T^A)_a^{\cdot\,b} |\psi_a \bar \psi^b\rangle \nn \\
& |\psi \psi\rangle^A_{{\rm{\bf S}}} = (T^A_{\rm S})_{ab} |\psi_a \psi _b\rangle  \ , 
\end{align}
where $\psi=\chi$ and/or $\eta$, $T^A$ are the $SU(N)$ generators, and $T^A_S$ are $N(N+1)/2$ symmetric matrices that we choose to be
the symmetric $SU(N)$ generators, with the addition of $\frac{\mathbb{1}_N}{\sqrt{2N}}$, which is normalized to preserve the canonical trace normalization ${\rm Tr}[T_S^A T_S^B]=\frac{\delta^{AB}}{2}$. Note that in this case the  symmetric combination is always built through two identical states: this is the reason of the extra factor $1/\sqrt 2$ in the symmetric two particle state with respect to the adjoint one. By  direct computation one obtains the following non zero group factor amplitudes relevant for the scattering in $J=0$, see again Eq.~\eqref{eq:full_matrix},
\begin{align}\label{eq:group_model1_J0}
++++
\begin{cases}
{\cal F}^{s,{\bf 1}}_{\bar \chi \eta \bar \chi \eta} = N
\end{cases}
\qquad \qquad
++--
\begin{cases}
{\cal F}^{s, \bf{1}}_{\bar \chi \eta  \chi \bar \eta} = N \\
{\cal F}^{t, \bf{1}}_{\bar \chi \eta  \chi \bar \eta} = 1 \\
{\cal F}^{t,{\rm Adj}}_{\bar \chi \eta  \chi \bar \eta} =1 \\
{\cal F}^{t,{\rm {\bf S}}}_{\eta\eta\chi\chi} = {\cal F}^{u,{\rm {\bf S}}}_{\eta\eta\chi\chi}  = \frac{1}{2} \\
\end{cases} \ ,
\end{align}
where the $----$ and $--++$ amplitudes are equivalent since they are obtained by conjugation.
Note that among the vanishing $++++$ group factors we have that, for example, the one in the adjoint channel is zero because $S$ is a scalar singlet and the amplitude thus turns out to be proportional to the trace of the $SU(N)$ generators, while the one in the symmetric channel is zero since the theory does not mediate processes such as $\eta\eta\to\eta\eta$. 
From the explicit form of the group factors it is clear that it's the singlet channel that manifests an enhancement of the scattering amplitude due to the $SU(N)$ group structure. By using the general expression
of Eq.~\eqref{eq:total_T}, the $J=0$ partial wave in the singlet channel written in the $(\bar\chi \eta, \chi\bar \eta)$ basis reads
\begin{align}
\label{eq:model1_j0}
a^{J=0}_{{\bf 1}}& =\frac{y^2}{32\pi} \int_{-1}^{+1}{\rm d}\cos\theta \; d^0_{00}(\theta)
\begin{pmatrix}
N {\cal T}^{++++}_s  & N {\cal T}^{++--}_s  + {\cal T}^{++--}_t \\
 N {\cal T}^{--++}_s  + {\cal T}^{--++}_t & N {\cal T}^{----}_s 
\end{pmatrix} = \nn \\
& =  \frac{ y^2 }{16\pi}
\begin{pmatrix}
- N & N+1 \\
N+1 & - N
\end{pmatrix} \ , 
\end{align}
whose largest eigenvalue in absolute value is $\frac{y^2}{16\pi}(2N+1)$. The eigenvalues relative to the scatterings in the other $SU(N)$ irreducible representations are all smaller and thus the PU condition of Eq.~\eqref{eq:pu-limit}  leads to the bound
\be
y^2 < \frac{8\pi}{2N+1}\ .
\ee
If the scalar is complex the conservation of the $U(1)$ charge forbids scattering in the $\pm\pm\mp\mp$ channel. The matrix of Eq.~\eqref{eq:model1_j0} becomes thus diagonal with eigenvalues $\frac{y^2}{16\pi} N$ and the bound now reads
\be\label{eq:model1_J0_complex}
y^2 < \frac{8 \pi}{N} \ ,
\ee
which is weaker than in the real scalar case. 

For half-integer $J$ the two-particle states with $+0$ helicities decompose for real scalar $S$ as
\be
+0
\begin{cases}
 \bar\chi S  \sim {\overline{{\tiny{\yng(1)}}}} \nn \\
 \eta S \sim {\tiny{\yng(1)}}
\end{cases}  \ ,
\ee
while the states with $-0$ helicities are their conjugates. The group factors read
\begin{align}\label{eq:group_model1_J12}
+0+0
\begin{cases}
{\cal F}^{s,{{\overline{\tiny{\yng(1)}}}}}_{\bar \chi S \bar \chi S} = {\cal F}^{u,{\overline{\tiny{\yng(1)}}}}_{\bar\chi S \bar\chi S} = {\cal F}^{s,{\tiny{\yng(1)}}}_{\eta S \eta S} = {\cal F}^{u,{\tiny{\yng(1)}}}_{\eta S \eta S} = 1 \ ,
\end{cases}
\end{align}
which are equivalent to their conjugates.
Since there is no scattering between the $+0$ and $-0$ states in the massless limit, we can consider only the $+0+0$ scattering channel, while the $-0-0$ will be its conjugate. 
For example for the former the scattering 
in the fundamental channel for $J=1/2$ reads
\be\label{eq:model1_j12}
a^{J=\frac{1}{2}}_{{\tiny{\yng(1)}}} =
-\frac{y^2}{32\pi}
 \int_{-1}^{+1}{\rm d}\cos\theta \;
({\cal T}_s^{+0+0} + {\cal T}_u^{+0+0})
d^\frac{1}{2}_{\frac{1}{2}\frac{1}{2}}(\theta)
 \, {\mathbb{1}}_N  = -  \frac{3}{32}\lambda^2 \,{\mathbb{1}}_N \ ,
\ee
thus leading to the bound
\be
y^2 < \frac{16\pi}{3} \ .
\ee
One can understand that there is no multiplicity factor due to the $SU(N)$ group structure since for all the diagrams, both in the $s-$ and $u-$channel, the $SU(N)$ index is never contracted between initial or final states, but is instead conserved between them.

If $S$ is a complex scalar  the $U(1)$ charge conservation enforces to treat separately the scattering of two distinct Mandelstam channels. As an example both the $\eta S \to \eta S$  and the $\eta S^* \to \eta S^*$ scattering proceed through the fundamental representation, but the former via an  $s-$channel diagram, while the latter via  $u-$channel one. The different angular function of the two amplitudes makes the one in the $u-$channel dominate.  For the scattering $\eta S^* \to \eta S^*$ one obtains
\be\label{eq:model1_j12_complex}
a^{J=\frac{1}{2}}_{{\overline{{\tiny{\yng(1)}}}}} =
-\frac{y^2}{32\pi}
 \int_{-1}^{+1}{\rm d}\cos\theta \;
{\cal T}_u^{+0+0}
d^\frac{1}{2}_{\frac{1}{2}\frac{1}{2}}(\theta)
\, {\mathbb{1}}_N  = -  \frac{y^2}{16\pi} \, {\mathbb{1}}_N  \ ,
\ee
which leads to the bound
\be
y^2 < 8 \pi \ .
\ee
For both the real and complex scalar the bounds in the $J=1/2$ sector are weaker than in the $J=0$ one.

\begin{figure}[t!]
\begin{center}
\includegraphics[width=0.55\textwidth]{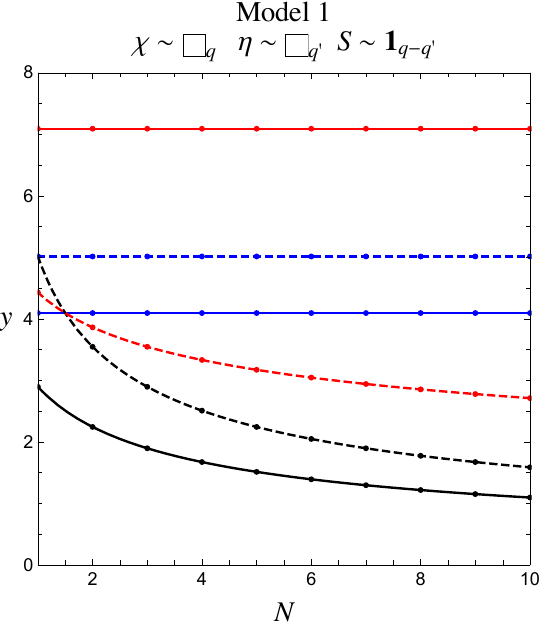}
\caption{\small PU bounds on the Yukawa couplings $y$ for the model 1 of the {\emph{Dirac type}} class
for $J=0$ (black), $J=1/2$ (blue) and $J=1$ (red).
The solid (dashed) lines correspond to the case of a real (complex) scalar field. }
\label{fig:model1}
\end{center}
\end{figure}

Finally we can consider the relevant scatterings for integer $J>0$, starting again with the case of a real scalar field.
Here the two particle states among fermions decompose as
\be\label{eq:model1_J1_dec}
+-
\begin{cases}
 \bar\chi \chi  \sim  {\bf 1}+ {\rm Adj} \nn \\
 \bar\chi \bar\eta  \sim{\rm{\bf{\overline{S}}}}+{\rm{\bf \overline{AS}}} \nn \\
\eta\chi  \sim  {\rm{\bf S}}+{\rm{\bf AS}} \nn \\
\eta\bar\eta \sim  {\bf 1}+ {\rm Adj} \nn \\
\end{cases}  \ ,
\ee
while, the scalar field being an $SU(N)$ singlet, only the trivial representation exists for $00$. Bose symmetry however forbids the scattering in the $00+-$ channel for odd $J$~\cite{Jacob:1959at}, while the relevant group factors in the $+-+-$ channel are\footnote{Note that the antisymmetric combination no longer vanishes here, since its built with two non identical fermions.}
\begin{align}\label{eq:group_model1_J1_A}
+-+-
\begin{cases}
{\cal F}^{u, \bf{1}}_{\bar \chi \chi  \eta \bar \eta} = {\cal F}^{u, \bf{1}}_{  \eta \bar \eta \bar \chi \chi} = 1 \\
{\cal F}^{u, {\rm Adj}}_{\bar \chi \chi  \eta \bar \eta} = {\cal F}^{u, \bf{1}}_{  \eta \bar \eta \bar \chi \chi} = 1 \\
{\cal F}^{u, {\bf S}}_{\bar \chi \chi  \eta \bar \eta} = {\cal F}^{u, \bf{1}}_{  \eta \bar \eta \bar \chi \chi} = 1 \\
{\cal F}^{u, \bf{AS}}_{\bar \chi \chi  \eta \bar \eta} = {\cal F}^{u, \bf{1}}_{  \eta \bar \eta \bar \chi \chi} = -1 \\
\end{cases}  \ .
\end{align}
For the lowest partial wave $J=1$ all the eigenvalues of the partial wave matrix are $\pm \frac{y^2}{32\pi}$, therefore the bound is simply
\be
y^2 < 16\pi \ .
\ee
In the case of a complex scalar field instead one has additional non vanishing processes as $S S^*\to  \bar\chi \chi  $ and $S S^*\to \eta \bar \eta$ that can however proceed only via the singlet channel, since the scalar belongs to the trivial $SU(N)$ representation. The scattering among singlet is thus modified with respect to the real scalar case. 
The group structure for these scatterings is
\begin{align}\label{eq:group_model1_J1_B}
00^*+-
\begin{cases}
{\cal F}^{u, \bf{1}}_{ S S^*\bar \chi \chi }  = {\cal F}^{u, \bf{1}}_{S S^*\eta\bar\eta  } = \sqrt N\\
\end{cases}  \ ,
\end{align}
and explicitly one has in the $(\bar\chi \chi, \eta\bar\eta, S S^*)$ basis
\begin{align}
\label{eq:model1_j1}
a^{J=1}_{{\bf 1}} & =
\frac{y^2}{32\pi}
 \int_{-1}^{+1}{\rm d}\cos\theta \;
\begin{pmatrix}
0 & d^1_{11}(\theta){\cal{T}}_u^{+-+-} & -d^1_{01}(\theta) {\cal{T}}_u^{00^*+-} \\
d^1_{11}(\theta){\cal{T}}_u^{+-+-}& 0 &  -d^1_{01}(\theta){\cal{T}}_u^{00^*+-} \\
 d^1_{10}(\theta){\cal{T}}_u^{00^*+-} &d^1_{10}(\theta) {\cal{T}}_u^{00^*+-}& 0 \\ 
\end{pmatrix}
\circ
\begin{pmatrix}
0 & 1 & \sqrt N \\
1 & 0 & \sqrt N \\
\sqrt N & \sqrt N & 0
\end{pmatrix} \nn = \\
& =- \frac{y^2}{32\pi}
\begin{pmatrix}
0 & 1 & \sqrt{2N} \\
1 & 0 &  \sqrt{2N} \\
 \sqrt{2N} &  \sqrt{2N} & 0  
\end{pmatrix} , 
\end{align}
where $\circ$ represents the Hadamard product\footnote{The Hadamard product $H$ of $n$ matrices  of the same dimension $H=A^1\circ \ldots \circ A^n$ is defined in components as $H_{ij}=A^1_{ij}\ldots A^n_{ij}$.}. Here we have split the Lorentz and group part of the amplitudes to highlight that the different helicity channels are associated to different group coefficients.
The largest eigenvalue in absolute value is $\frac{y^2}{64\pi}(1+\sqrt{1+16 N})$ and  the bound thus reads
\be\label{eq:model1_J1_complex}
y^2 < \frac{32 \pi}{1+\sqrt{1+16 N}} \ .
\ee
We then report in Fig.~\ref{fig:model1} the bounds on $y$ obtained in the $J=0,1/2$ and 1 partial waves. In particular we see that in the case of a real scalar field the strongest bound is obtained in the $J=0$ channel.
On the other hand for a complex scalar field the strongest bound is obtained through the analysis of the scattering in the $J=1$ channel, while the one in the $J=0$ channel dominates only in the case of an abelian theory. This is a non trivial result, highlighting the role that higher partial waves can have in deriving a PU bound.

%%%%%%%%%%%%%%%%%
%%%%%%%%%%%%%%%%%
%%%%%%%%%%%%%%%%%

\subsection{Second model: $\chi\sim {\tiny{\fbox{$\protect\phantom s$} }}_q$, $\eta\sim \mathbf{1}_{q^\prime}$, $S\sim {\tiny{\fbox{$\protect\phantom s$} }}_{q-q^\prime}$  }\label{sec:model2}

\begin{figure}[t!]
\begin{center}
\includegraphics[width=0.55\textwidth]{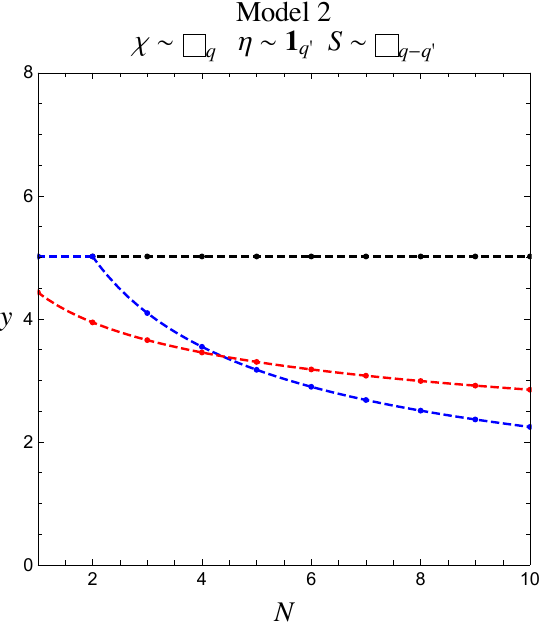} 
\caption{\small PU bounds on the Yukawa couplings $y$ for the model 2 of the {\emph{Dirac type}} class
for $J=0$ (black), $J=1/2$ (blue) and $J=1$ (red).
In this case the scalar is always a complex field. }
\label{fig:model2}
\end{center}
\end{figure}

In this  model the scalar transforms in the fundamental representation of $SU(N)$, and is thus always a complex field.  By fixing the basis
\be
\psi_L = (\chi_{a}, \eta^{c})^T \ , \qquad \phi= (S_a, S^{*a})^T \ ,
\ee
where $a$ runs from 1 to $N$, one has for the Yukawa matrix of Eq.~\eqref{eq:gen_lag}
\be
{\cal Y}_{\alpha i j} = 
y
\begin{cases}
\delta_{i,\alpha}\delta_{j,N+1} + (i \leftrightarrow j) \quad \alpha \le N\\
 \delta_{i,\alpha-N}\delta_{j,N+1} + (i \leftrightarrow j) \quad \alpha > N\\ 
\end{cases} \ ,
\ee
with $i,j=1,\dots,N+1$.
We start again by considering the $J=0$ partial wave. Here the two particle states decompose as
\be
++
\begin{cases}
 \bar\chi \eta  \sim  {\overline{{\tiny{\yng(1)}}}}\nn \\
 \bar\chi \bar \chi \sim  {\overline{{\rm{\bf S}}}}+{\overline{{\rm{\bf AS}}}} \\
 \eta \eta \sim {\bf 1}
\end{cases} \qquad
\ee 
and the $--$ states are the conjugates. Again, the antisymmetric combination identically vanishes for even $J$. The two-particle states can be written in analogy with Eq.~\eqref{eq:states}, where clearly the singlet combination is now the trivial state $|\eta \eta \rangle$, to which we add the state in the antifundamental as
\be
|\bar \chi \eta \rangle^a_{{\overline{{\tiny{\yng(1)}}}}} = | \bar\chi^a \eta  \rangle \ .
\ee
In this theory all the scatterings in the $\pm\pm \mp \mp$ channels vanish together with the ones that proceed through the singlet and symmetric channels in $\pm\pm\pm\pm$. The only non vanishing amplitudes are the ones in the (anti)fundamental channel with a group factor that reads
\begin{align}\label{eq:group_model2_J0}
++++
\begin{cases}
{\cal F}^{s,{\tiny{\yng(1)}}}_{\bar \chi \eta \bar \chi \eta}  = {\cal F}^{s,{{\overline{\tiny{\yng(1)}}}}}_{ \chi \bar\eta  \chi \bar\eta}= 1
\end{cases} \ .
\end{align}
In this case the partial wave matrix is already diagonal and after the trivial integration over the angular variable one obtains the bound
\be
y^2 < 8 \pi \ .
\ee
Moving onto the $J=1/2$ channel we can consider the $+0+0$ scattering. Here we have
 \be
+0
\begin{cases}
\eta S  \sim  {\tiny{\yng(1)}}\nn \\
\eta S^*  \sim  \overline{{\tiny{\yng(1)}}}\nn \\
\bar\chi S  \sim  {\bf 1}+{\rm Adj}  \nn \\
\bar\chi S^*  \sim {\rm{\bf{\overline{S}}}}+{\rm{\bf \overline{AS}}} \nn \\
\end{cases}
\ee
 and the group factors for the non zero amplitudes are
\begin{align}\label{eq:group_model2_J12}
+0+0
\begin{cases}
{\cal F}^{s,\mathbf{1}}_{\bar \chi S \bar \chi S}  = N \\
{\cal F}^{s,{\tiny{\yng(1)}}}_{\eta S \eta S} = {\cal F}^{u,{{\overline{\tiny{\yng(1)}}}}}_{\eta S \eta S} = 1 \\
{\cal F}^{u,{\rm{\bf{\overline{S}}}}}_{\bar\chi S^* \bar\chi S^*} = - {\cal F}^{u,{\rm{\bf{\overline{AS}}}}}_{\bar\chi S^*\bar\chi S^*} = 1\\
\end{cases} \ ,
\end{align}
with again the $-0-0$ being the conjugates.
We can consider the singlet channel, which exhibits a $N$ multiplicity factor. In the $+0+0$ sector one has
\be\label{eq:model2_j12}
a^{J=\frac{1}{2}}_{{\bf 1}} =
\frac{y^2}{32\pi}
 \int_{-1}^{+1}{\rm d}\cos\theta \;
{\cal T}_s^{+0+0}
  d^\frac{1}{2}_{\frac{1}{2}\frac{1}{2}}(\theta)
N
= -\frac{y^2}{32\pi} N \ . 
\ee
This is the largest eigenvalue for $N>1$ while for $N=1$ it's the scattering in the (anti)symmetric channels that dominates due to the $u-$channel amplitude which scales as ${\cal T}_u^{+0+0} \sim \frac{1}{\cos\frac{\theta}{2}}$ and has eigenvalue $\pm 2 y^2$. Altogether we obtain
\be\label{eq:model2_J12_complex}
y^2 <
\begin{cases}
8\pi \quad N=1 \\
\frac{16\pi}{N} \quad N\ge 2
\end{cases} \ .
\ee
Finally for $J=1$ the relevant two-particle states decompose as
 \be
+-
\begin{cases}
\bar\chi \chi \sim {\bf 1}+ {\rm Adj} \nn \\
\bar\eta \eta \sim {\bf 1} \nn \\
\eta \chi   \sim  {\tiny{\yng(1)}}\nn \\
\bar\chi \bar\eta   \sim  {\overline{{\tiny{\yng(1)}}}} \\
\end{cases}
\qquad
00
\begin{cases}
S S \sim {\rm{\bf{{S}}}}+{\rm{\bf {AS}}} \nn \\
S S^* \sim {\bf 1}+{\rm Adj} \nn \\
S^* S^* \sim {\rm{\bf{\overline{S}}}}+{\rm{\bf \overline{AS}}}
\end{cases} \ ,
\ee
where however now it's the symmetric combinations of the two identical scalars that vanishes identically for $J=1$~\cite{Jacob:1959at}. The group factors for the non-zero amplitudes are
\begin{align}\label{eq:group_model2_J1}
+-+-
\begin{cases}
{\cal F}^{u,{\bf 1}}_{\bar \chi \chi \eta \bar \eta} = {\cal F}^{u,{\bf 1}}_{\eta \bar \eta \bar \chi \chi}  = \sqrt N \\
{\cal F}^{u,{\tiny{\yng(1)}}}_{\eta\chi\eta\chi}  = {\cal F}^{u,{\tiny{\yng(1)}}}_{\bar\chi\bar\eta\bar\chi\bar\eta} = 1
\end{cases}
\qquad \qquad
00^*+-
\begin{cases}
{\cal F}^{u,{\bf 1}}_{ S S^* \bar \chi \chi} = 1 \\
{\cal F}^{u,{\bf 1}}_{S S^*\eta \bar \eta } = \sqrt N \\
{\cal F}^{u,{\rm Adj}}_{S S^*\bar\chi \chi } =1\\
\end{cases} \ .
\end{align}
Note that the $00^*+-$ scattering in the singlet channel has two contributions, coming from the two possible ways of making an $SU(N)$ singlet from the two fermions. The strongest bound turns out again to be the one arising from the scattering among singlets. Explicitly one has in the $(\bar\chi\chi, \eta\bar \eta,S S^*)$ basis
\begin{align}
\label{eq:model2_j1}
a^{J=1}_{{\bf 1}} & =
\frac{y^2}{32\pi}
 \int_{-1}^{+1}{\rm d}\cos\theta \;
\begin{pmatrix}
0 & d^1_{11}(\theta){\cal{T}}_u^{+-+-} & -d^1_{01}(\theta) {\cal{T}}_u^{00^*+-} \\
d^1_{11}(\theta){\cal{T}}_u^{+-+-}& 0 &  -d^1_{01}(\theta){\cal{T}}_u^{00^*+-} \\
 d^1_{10}(\theta){\cal{T}}_u^{00^*+-} &d^1_{10}(\theta) {\cal{T}}_u^{00^*+-}& 0 \\ 
\end{pmatrix}
\circ
\begin{pmatrix}
0 & \sqrt N & 1 \\
\sqrt N& 0 & \sqrt N \\
1 & \sqrt N & 0
\end{pmatrix}\nn \\
& =-\frac{y^2}{32\pi}
\begin{pmatrix}
0 & \sqrt N & \sqrt{2} \\
\sqrt N & 0 & \sqrt{2N} \\
\sqrt{2}&  \sqrt{2N} & 0
\end{pmatrix} ,
\end{align}
where we again have split for convenience the Lorentz and group structure of the amplitude. The eigenvalues of this matrix have a complicated form for generic $N$ and we thus show the numerical results in Fig.~\ref{fig:model2}. There we see that for $N<4$ it's the $J=1$ partial wave that enforces the strongest bound while for $N\ge 5$ is the $J=\frac{1}{2}$ one.  As for the case of the previous toy model we see that the stronger limit can arise from partial waves different from $J=0$. We also note that when $N=1$ we recover, with no ambiguity, the same bounds obtained for the first toy model in the case of a complex scalar field.

\section{The case of the Standard Model Yukawa sector}\label{sec:sm_yuk}

The two toy models discussed in the previous Section can be used as building blocks through which it is possible to study more involved theories where, {\emph{e.g.}}, the fields are charged under multiple $SU(N_i)$ factors and/or where more than one generation of fields with the same quantum numbers is present. We highlight this by discussing in detail the case of the SM Yukawa couplings, focusing on the down type quark sector, for which the Yukawa Lagrangian is
\be\label{eq:sm_yuk}
-{\cal L} = y_d^{ij} \bar q^i H d^j + h.c. \ ,
\ee
where $q^i$ and $d^j$ are left-handed and right-handed fermions field respectively and where the flavor indices $i$ and $j$ run from 1 to 3. From the point of view of the gauge symmetries,
this theory belongs to the first {\emph{Dirac type}} model described in Sec.~\ref{sec:model1}  with respect to $SU(3)_c$ and to the second  {\emph{Dirac type}} model described in Sec.~\ref{sec:model2} with respect to $SU(2)_L$, with $H$ being a complex scalar field.

\subsection{Multiple $SU(N_i)$ factors}\label{sec:sm_gauge_mult}

We start by discussing the role played by multiple $SU(N_i)$ factors under which the various fields can be charged. To this end we consider a single generation of SM fermions, $y_d^{11}=y_d$. The rule of Eq.~\eqref{eq:total_T} is readily generalized, by considering that now each group  factor coefficient ${\cal F}$ is the product of the various group coefficient factors for the different $SU(N)$ groups and the dimension of the identity matrix is the product of the dimensions of the considered irreducible representations for each $SU(N)$ factor. It is again instructive to work out the most important scattering amplitudes for the case of $J=0$, half-integer $J$ and integer $J>0$. We start  with $J=0$. Since the scalar is complex the only non-zero amplitudes are in the $\pm\pm\pm\pm$ scattering channel. Working in the $(\bar q d, q \bar d)$ basis one has that the scattering proceeds through the (anti)fundamental channel for what concerns $SU(2)_L$ with a group factor proportional to the identity, see Eq.~\eqref{eq:group_model2_J0}.
As regarding $SU(3)_c$ one can again consider the scattering in the singlet channel, which exhibits a group factor enhancement, see Eq.~\eqref{eq:group_model1_J0}.
The $J=0$ partial wave explicitly reads
\be
a^{J=0}_{SU(3)={\bf 1}, SU(2)= {\tiny{\yng(1)}}}= \frac{y_d^2}{32\pi} \int_{-1}^{+1}{\rm d}\cos\theta d^0_{00}(\theta) 
\begin{pmatrix}
N_3 {\cal T}_s^{++++}  & 0 \\
0 &N_3 {\cal T}_s^{----} 
\end{pmatrix}
\times
 {\mathbb 1}_{N_2} \ ,
\ee 
where $N_3=3$, $N_2=2$.
 Here the  presence of two $SU(N)$ groups simply increases the eigenvalue multiplicity, given that $SU(2)_L$ group factor is proportional to the identity and the scattering matrix is already diagonal. The bound in this case is simply the one of Eq.~\eqref{eq:model1_J0_complex}
\be\label{eq:bound_sm_yuk}
y_d < \sqrt{\frac{8\pi}{N_3}} = \sqrt{\frac{8\pi}{3}} \sim 2.9 \ .
\ee
For $J=1/2$ we can consider the $+0+0$ sector and it's convenient to consider the $\bar q H \leftrightarrow \bar q H$ in the antifundamental channel for $SU(3)$ and singlet channel for $SU(2)$. The group factors
can be read from Eq.~\eqref{eq:group_model1_J12} and Eq.~\eqref{eq:group_model2_J12} and are the identity for $SU(3)$ and $N_2=2$ for  $SU(2)$. Also in this case the presence of two $SU(N)$ factors simply increases the eigenvalue multiplicity.  The partial wave reads
\be
a^{J=1/2}_{SU(3)= {{\overline{\tiny{\yng(1)}}}},SU(2)={\bf 1}}= \frac{y_d^2}{32\pi} \int_{-1}^{+1}{\rm d}\cos\theta d^\frac{1}{2}_{\frac{1}{2}\frac{1}{2}}(\theta) {\cal T}_s^{+0+0}
{\mathbb 1}_{N_3} N_2 = -\frac{y^2}{32\pi} {\mathbb 1}_{N_3} N_2 \ ,
\ee 
leading to the same bound of Eq.~\eqref{eq:model2_J12_complex}
\be
y_d < \sqrt{\frac{16\pi}{N_2}} = \sqrt{8 \pi} \sim 5 \ .
\ee
The situation is more involved for $J=1$. Focusing on the singlet channel scattering for both the $SU(3)_c$ and $SU(2)_L$ groups one has the group coefficients of Eq.~\eqref{eq:group_model1_J1_A}, Eq.~\eqref{eq:group_model1_J1_B} and Eq.~\eqref{eq:group_model2_J1}.
Explicitly then in the $(\bar q_L q_L, d_R \bar d_R, H H^*)$ basis one obtains
\begin{align}
\label{eq:modelSM_j1}
a^{J=1}_{{\bf 1}} & =
\frac{y_d^2}{32\pi}
 \int_{-1}^{+1}{\rm d}\cos\theta \;
\begin{pmatrix}
0 & d^1_{11}(\theta){\cal{T}}_u^{+-+-} & -d^1_{01}(\theta) {\cal{T}}_u^{00^*+-} \\
d^1_{11}(\theta){\cal{T}}_u^{+-+-}& 0 &  -d^1_{01}(\theta){\cal{T}}_u^{00^*+-} \\
 d^1_{10}(\theta){\cal{T}}_u^{00^*+-} &d^1_{10}(\theta) {\cal{T}}_u^{00^*+-}& 0 \\ 
\end{pmatrix}
\circ
\begin{pmatrix}
0 & 1 & \sqrt{N_3} \\
1 & 0 &  \sqrt{N_3} \\
 \sqrt{N_3} &  \sqrt{N_3} & 0  
\end{pmatrix} \nn \circ \\
& \circ  
\begin{pmatrix}
0 & \sqrt{N_2} & \sqrt{2} \\
\sqrt{N_2} & 0 & \sqrt{2N_2} \\
\sqrt{2}&  \sqrt{2N_2} & 0
\end{pmatrix}  
 =- \frac{y^2}{32\pi}
\begin{pmatrix}
0 & \sqrt{N_2} & \sqrt{2 N_2 N_3} \\
\sqrt{N_2} & 0 & \sqrt{2 N_3} \\
\sqrt{2 N_2 N_3} & \sqrt{2 N_3} & 0 
\end{pmatrix} , 
\end{align}
which leads to the bound
\be
y_d \lesssim 3.2 \ .
\ee
Here we see the non trivial interplay between the two $SU(N)$ factors, which for this partial wave gives a bound which is stronger than the one obtained considering only one of the two $SU(N)$ factors, see Eq.~\eqref{eq:model1_J1_complex} and Eq.~\eqref{eq:model2_j1}.
Overall in the case of the SM Yukawa sector the most stringent bound turns out then to arise from $J=0$ partial wave. The limit of Eq.~\eqref{eq:bound_sm_yuk} implies that if there were additional quarks acquiring mass from EWSB, their mass should have been $\lesssim 500\;$GeV
in order to preserve PU. Analogously,  additional leptons should have a mass $\lesssim 700\;$GeV.

\subsection{Multiple generations}

We want now to highlight what is the role played by the presence of mutiple states with the same quantum numbers, as in the case of multiple generations of SM fermions. We then go back the the general case of Eq.~\eqref{eq:sm_yuk} and, for simplicity, work with only two generations of fermions. Through a biunitary rotation acting on the fermion fields $q_L \to U_L q_L$ and  $d_R = U_R d_R$ it is possible to go to a basis where the Yukawa matrix becomes diagonal with real and non negative entries, namely
\be
\tilde y_d = U_L y_d U_R^\dag \ .
\ee
It's interesting then to ask what happens to the largest eigenvalues obtained in the single family case of Sec.~\ref{sec:sm_gauge_mult}. In the $J=0$ sector the Higgs boson can mediate $s-$channel scatterings among different generations. Choosing as basis $(\bar q_1 d_1,\bar q_2 d_2, q_1 \bar d_1, q_2 \bar d_2)$ the partial wave matrix in the singlet channel becomes, after the angular integration,
\be
a^{J=0}_{SU(3)={\bf 1}, SU(2)= {\tiny{\yng(1)}}}=\frac{ N_3}{16\pi}
\begin{pmatrix}
\tilde{y}_{d,1}^2 & \tilde{y}_{d,1} \tilde{y}_{d,2} & & \\
  \tilde{y}_{d,1} \tilde{y}_{d,2} & \tilde{y}_{d,2}^2& & \\
& & \tilde{y}_{d,1}^2 & \tilde{y}_{d,1} \tilde{y}_{d,2}  \\
& & \tilde{y}_{d,1} \tilde{y}_{d,2} & \tilde{y}_{d,1}^2 &   \\
\end{pmatrix}
\times
 {\mathbb 1}_{N_2} \ ,
\ee
whose largest eigenvalue is $\frac{N_3}{16\pi} (\tilde{y}_{d,1}^2+\tilde{y}_{d,2}^2)$ and the bound is thus on the geometric mean of the two Yukawa couplings
\be
\sqrt{\tilde y_{d,1}^2+\tilde y_{d,2}^2} < \sqrt{\frac{8\pi}{N_3}} = \sqrt{\frac{8\pi}{3}} \sim 2.9 \ .
\ee
This is not the case for the scattering in $J=1/2$, where different generations do not communicate since the scatterings proceed through the exchange of an $s-$ or $u-$channel fermion. 
In this case the eigenvalues give independent bounds on the two couplings separately, which read 
\be
\label{eq:2generations}
\tilde y_{d,1},\tilde y_{d,2} < \sqrt{\frac{16\pi}{N_2}} = \sqrt{8 \pi} \sim 5 \ ,
\ee
in total analogy with the single family case. In $J=1$ the situation is again more involved due to the presence of the $00$ two particle state which
is common between all generations. The eigenvalues of the scattering matrix have a complicated analytical form, but numerically one can see that the strongest bound is always given by the $J=0$ partial wave.

\section{Majorana type theories}\label{sec:gauge-majo}

In this Section we study the {\emph{Majorana type}} theories, described by Eq.~\eqref{eq:model_II}.
As discussed in Sec.~\ref{sec:toy} and indicated in Tab.~\ref{tab:models}, when only one $SU(N)$ factor is present, or flavor multiplicity is not added, only one model, other than the one where all fields are $SU(N)$ singlets, can be written. We firstly study these two theories in turn and then in Sec.~\ref{sec:qq} we present an explicit, phenomenologically relevant, example in which the involved fields are charged under multiple $SU(N)$ factors and the above caveat can thus be evaded. Given that the two models of Tab.~\ref{tab:models} cannot be written for arbitrary $SU(N)$, in this Section we directly illustrate our findings without explicitly presenting the group factors ${\cal F}$, as opposed to the thorough derivation of Sec.~\ref{sec:gauge-dirac} for the {\emph{Dirac type}} theories. We also do the same for the explicit example of Sec.~\ref{sec:qq}. In this case the various amplitudes can be derived analogously to the examples of Sec.~\ref{sec:gauge-dirac}.

\subsection{First model: $\chi\sim {\mathbf{1}}_q$, $S \sim {\mathbf{1}}_{2q}$}

In this model both the fermion and the scalar are $SU(N)$ singlets, where the latter is real if $q=0$. Clearly, no group factor is present in this theory and the partial waves can be easily built directly from the Lorentz amplitudes of Tab.~\ref{tab:lorentz_majorana}. In the complex scalar basis the Yukawa matrix of Eq.~\eqref{eq:gen_lag} is simply ${\cal Y}_\alpha=y$, where $\alpha=1$ in the real scalar case and $\alpha=1,2$ in the complex scalar one. Let's start again by discussing $J=0$ with real $S$. In this case both the $\pm\pm\pm\pm$ and $\pm\pm\mp\mp$ helicity channels contribute. The partial wave matrix is readily computed and, after integration in the $(\bar \chi,\chi)$ basis, reads
\begin{align}
\label{eq:majo1}
a^{J=0}& =\frac{y^2}{32\pi} 
\begin{pmatrix}
-1 & 3 \\
3 & -1
\end{pmatrix} \ ,
\end{align}
which gives the bound
\be
y^2 < 4\pi \ .
\ee
Again, if the scalar is complex there is no scattering in the $\pm\pm\mp\mp$ helicity channels. The partial wave matrix of Eq.~\eqref{eq:majo1} becomes diagonal and the bound relaxes to
\be
y^2 < 16\pi \ .
\ee
Moving to the $J=1/2$ partial wave, here we have a situation completely analogous to the one of Sec.~\ref{sec:model1} and the inferred bound are thus
\be
y^2 <
\begin{cases}
\frac{16\pi}{3} \qquad {\rm Real}~$S$ \\
8\pi \qquad \,\,{\rm Complex}~$S$
\end{cases} \ .
\ee
For $J=1$ in the real scalar case we have again only contributions from the $+-+-$ helicity channel and again we are in a configuration analogous to the one of  Sec.~\ref{sec:model1}. The bound can be directly read from the Lorentz part of the scattering amplitude and reads
\be
y^2 < 16\pi \ .
\ee
If $S$ is a complex scalar there is now a contribution to the partial wave matrix from the $00^*+-$ scatterings. In the basis $(SS^*,\bar\chi\chi)$ and after the angular integration the partial wave matrix is
\begin{align}
\label{eq:majo1_J1}
a^{J=1}& =\frac{y^2}{32\pi} 
\begin{pmatrix}
0 & -\sqrt 2 \\
-\sqrt 2 & -1
\end{pmatrix} \ ,
\end{align} 
which gives the bound
\be
y^2 < 8\pi \ .
\ee

\subsection{Second model: $\chi\sim {\tiny{\fbox{$\protect\phantom s$} }}_q$, $S\sim {\rm Adj}_{2q}$, $N=2$}

In this model the fermion transforms in the fundamental of $SU(2)$ while $S$ in the Adjoint representation, and is then a real scalar if $q=0$ and complex otherwise. In the first case we choose as basis
\be
\psi_L = (\chi_L)^T \ , \quad \phi = (S^A)^T \ , \quad A=1,2,3 \ ,
\ee
and the Yukawa matrix is then
\be
{\cal Y}_{\alpha i j} = y (T^\alpha\varepsilon)_{i}^{\cdot j} \ ,
\ee
where $\alpha=1,2,3$ and $i,j=1,2$.
When $q\ne 0$ then $S$ is a complex field. In this case we can choose
\be
\phi= (S^A, S^{A*})^T  \ , \quad A=1,2,3 \ ,
\ee
and the Yukawa is now
\be
{\cal Y}_{\alpha i j} =y
\begin{cases}
(T^\alpha\varepsilon)_{i}^{\cdot j} \ , \quad \alpha \le 3 \\
(T^{\alpha-3}\varepsilon)_{i}^{\cdot j} \ , \quad \alpha > 3
\end{cases} \ ,
\ee
where now $\alpha=1,\dots,6$.
Proceeding in a similar manner as for the Dirac type models, we find that in $J=0$ the bound is the same for both real and complex scalar, due to a cancellation between the $s$-, $t$- and $u$-channels in the $\pm\pm\mp\mp$ amplitudes. Moreover, since for the $\pm\pm\pm\pm$ transition there is only the $s$-channel exchange of $S$, the only non-vanishing scattering has the fermions in the $SU(2)$ triplet configuration, giving
\be
	y^2 < 32 \pi \,.
\ee
Moving to the $J=1/2$ partial wave, again the strongest bound is the same for real and complex scalars, coming from the scattering in the $\mathbf{4}$ of $SU(2)$:
\be
	y^2 < 16\pi \,.
\ee 
Finally, in $J=1$, the best bound for real $S$ is obtained in the adjoint channel, where the partial wave matrix reads
\be
	a^{J=1}_{\rm{Adj}} = \frac{y^2}{32\pi}
	\begin{pmatrix}
		-\frac{1}{4} & -i \\
		i & 0
	\end{pmatrix}
	\times \mathbb{1}_3 \,,
\ee
with eigenvalues $-\frac{y^2}{256\pi}(1\pm \sqrt{65})$, thus giving the bound 
\be
	y^2 < \frac{128 \pi}{1+\sqrt{65}} \,.
\ee
For complex $S$, instead, the singlet channel gives the strongest constraint. The partial wave matrix is
\be
	a^{J=1}_{\bf{1}} = \frac{y^2}{32\pi}
	\begin{pmatrix}
		-\frac{3}{4} & -\frac{\sqrt{3}}{2} \\
		-\frac{\sqrt{3}}{2} & 0
	\end{pmatrix}
	 \,,
\ee
and it has eigenvalues $-\frac{y^2}{256\pi}(3 \pm \sqrt{57})$. The bound therefore is
\be
	y^2 < \frac{128 \pi}{3+\sqrt{57}} \,.
\ee

\subsection{The case of the $S_1$ leptoquark}\label{sec:qq}

As already mentioned above, there are more possibilities for the Majorana type models once one allows for the fields to be charged under more than one $SU(N)$  group. Of particular phenomenological interest is the case of the leptoquark $S_1$, that will be discussed in more detail also in Sec.~\ref{sec:pheno}. This field transforms under the SM gauge group as $S_1 \sim ({\bf \bar 3}, {\bf 1}, \frac{1}{3})$.
One can thus write the following interaction term with the SM quark doublet
\be
	-{\cal L} = \frac{1}{2} y S_1 {\bar q}_L q^c_L + h.c. \,,
\ee
where the colour indices are contracted with the totally antisymmetric tensor $\varepsilon^{abc}$ of $SU(3)$, compensating the $SU(2)$ contraction $\bar q \varepsilon q^c$.
The bounds on the coupling $y$ in this case can be obtained along the same lines as the ones in the previous sections, and we therefore quote only the results for the three considered partial waves
\be
y^2 < 
\begin{cases}
4\pi \,\,\,\quad\quad J=0 \\
4\pi \,\,\,\quad\quad J=\frac{1}{2} \\
\frac{16\pi}{1 + \sqrt{17}} \quad J = 1
\end{cases} \,.
\ee

%%%%%%%%%%%%%%%%%%
%%% 	PHENO		%%
%%%%%%%%%%%%%%%%%%

\section{Phenomenological applications}\label{sec:pheno}

In this Section we apply our results to some illustrative models which present additional Yukawa interactions formulated to solve several anomalies reported in low energy measurements, such as the muon anomalous magnetic moment $(g-2)_\mu$ and the anomalies in the charged- and neutral-current decays of $B-$mesons, commonly dubbed as $R_{D^{(*)}}$ and $R_{K^{(*)}}$ anomalies respectively. The former is an anomaly in the ${\cal B}(B\to D^{(*)} \tau \nu)/{\cal B}(B\to D^{(*)} \ell \nu)$
observable in $b\to c\tau \nu$ charged-current transitions, with $\ell=e,\mu$, while the latter is an anomaly in the ${\cal B}(B\to K^{(*)} \mu^+\mu^-)/{\cal B}(B\to K^{(*)} e^+e^-)$ observable in $b\to s \mu^+\mu^-$ neutral-current transitions.  In order to explain the $R_{D^{(*)}}$ anomaly, a  $\sim 15\%$ modification with respect to the theory prediction is required. However in the SM the partonic process $b\to c\tau \nu$ occurs at tree-level, hence when one tries to explain the experimental measured value through some additional NP contribution one might encounter several problems. Since the NP contribution to this observable  scales, in case of a tree-level effect, as
\be
\delta R_{D^{(*)}} \sim \dfrac{g_{{\rm NP}}^2}{m^2_{\rm NP}}  \ ,
\ee
where $g_{{\rm NP}}$ and $m_{{\rm NP}}$ are the coupling and the mass of the relevant NP state, a large effect can be obtained either with a small NP mass or with a large NP coupling. However given that the suppression scale for the SM effective operator  $\frac{1}{\Lambda^2}(\bar q_2 \gamma^\mu \sigma^A q_3)(\ell_3 \gamma_\mu \sigma^a \ell_3)$ that can address this anomaly is $\Lambda\simeq 3\;$TeV, in the former case one has to face stringent limits from direct searches from, {\emph{e.g.}}, the LHC, while in the latter case the coupling might be pushed at the edge of perturbativity.
On the other hand the partonic process $b\to s\mu^+\mu^-$ entering the $R_{K^{(*)}}$ anomaly occurs in the SM at one-loop level, with a $V_{tb}V_{ts}$ CKM suppression. When considering NP models that try to explain this measurement also at one-loop level, again one can obtain couplings which might be in conflict with the requirement of perturbative unitarity.
The purpose of this Section is to apply our results to phenomenologically relevant models and show that the requirement of PU can enforce significant bounds that might deserve further investigation.
 Since typically in the models that we will consider more than two couplings at the same time can enter the expression of the PU limit, our strategy will be to trade some of them for other measurements and/or constraints and then to depict the PU bound in the region of the two remaining independent couplings.

 %%%%%%%%%%%%%%%%%%%%%%%%%%%%%
%%%%%%%%%%%%%%%%%%%%%%%%%%%%%

\subsection{Scalars and fermions for $R_{K^{(*)}}$ and $(g-2)_\mu$ anomalies}

The first model that we study extends the SM by adding new scalars and fermions in order to generate contributions to $b\to s \mu^+ \mu^-$ and $(g-2)_\mu$, both at loop-level  and it is based on~\cite{Gripaios:2015gra,Arnan:2016cpy}. We first consider the simplest extension which contains only left-handed (LH) couplings and then we evaluate the consequences of adding right-handed (RH) couplings.
\subsubsection{Left-handed scenario}\label{eq:LH}
In the LH scenario, the NP states couple only to LH SM quarks and leptons. We can consider two models with the following schematic interactions\footnote{Note that one can also construct a model where $\Phi$ and $\Psi$ couple to SM quarks while the conjugate fields $\Phi^c$ and $\Psi^c$ couple to leptons. This however leads to very similar phenomenological results.}\begin{itemize}
\item Model a) with one additional scalar $\Phi$ and two additional fermions $\Psi_q$ and $\Psi_\ell$ 
\begin{eqnarray}
-{\cal L}_{a} =  {\lambda^q_i\bar \Psi_q }{q^i_L}{\Phi} + \lambda^\ell_i \bar \Psi_\ell {\ell_L^i }{\Phi} +
h.c. \, \ . 
\label{eq:Linta}
\end{eqnarray}
\item Model b) with two additional scalars $\Phi_q$ and $\Phi_\ell$ and one additional fermion $\Psi$ 
\begin{eqnarray}
-{\cal L}_{b}=  {\lambda^q_i\bar \Psi }{q_L^i}{\Phi_q} + \lambda^\ell_i \bar \Psi {\ell_L^i}{\Phi_\ell} +
h.c. \, \ ,
\label{eq:Lintb}
\end{eqnarray}
\end{itemize}
where $q$ and $\ell$ are the SM quark and lepton doublet respectively and where the NP fields quantum numbers under the SM gauge group are at this level unspecified. Here however we wish to assess how constraining the PU requirement could be and since, as shown in Sec.~\ref{sec:gauge-dirac}, bounds are generally stringent when the theory features a real scalar field, we wish to consider models that feature a real scalar. In model a), however, by making $\Phi$ a real scalar one obtains an exact cancellation of the various contributions to $b\to s\mu^+\mu^-$~\cite{Arnan:2019uhr}, an option disfavored if one is willing to explain the $R_{K^{(*)}}$ anomaly. This is not the case for model b), where one can choose $\Phi_\ell$ to be a real scalar.

\begin{figure}[t!]
\begin{center}
\includegraphics[width=0.95\textwidth]{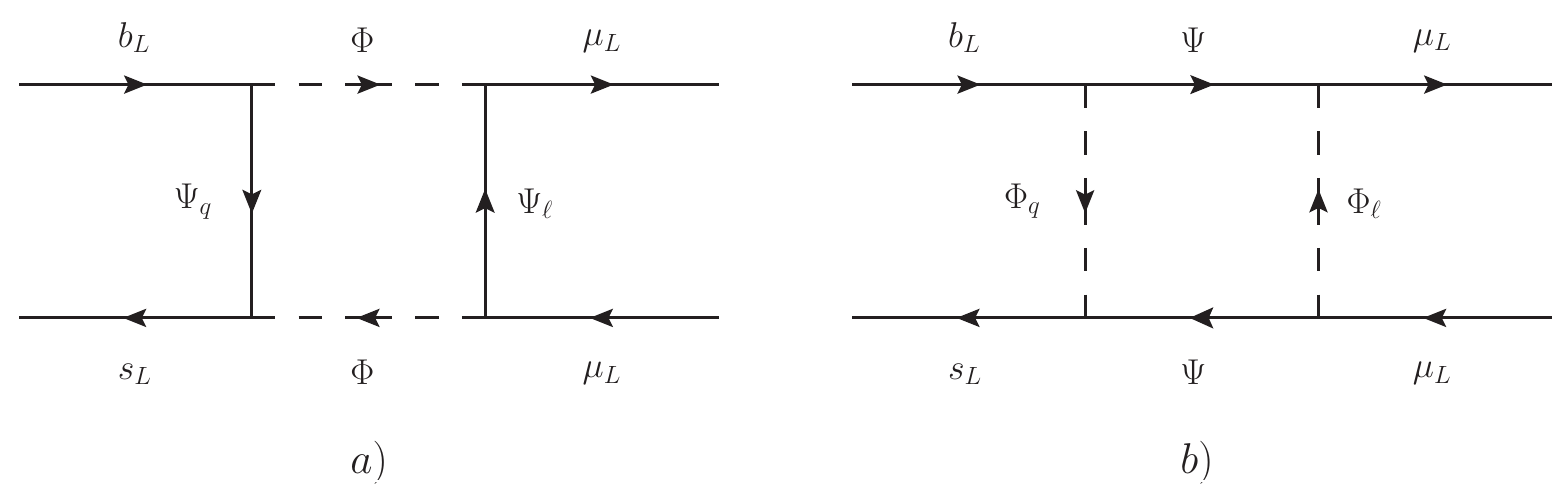}  
\caption{\small Loop-level diagrams responsible for generating the NP contribution to $b\to s\mu^+\mu^-$ for the models of Eq.~\eqref{eq:Linta} and Eq.~\eqref{eq:Lintb}.}
\label{fig:pheno_box}
\end{center}
\end{figure}

 Altogether we consider the following quantum number assignments under the SM gauge group for the two models
\begin{equation}
\begin{array}{cccc}
\text{model a)}~~& \Phi\sim ({\bf 1},{\bf 1},X)~~& \Psi_\ell \sim ({\bf 1},{\bf 1},-\frac{1}{2}+X)~~&\Psi_q \sim ({\bf 3},{\bf 1},\frac{1}{6}+X)\,, \\
\text{model b)}~~&\Psi\sim ({\bf 1},{\bf 2},X)~~&\Phi_\ell \sim ({\bf 1},{\bf 1},-\frac{1}{2}+X)~~&\Phi_q \sim ({\bf 3},{\bf 1},\frac{1}{6}+X) \, \ ,
\end{array}
\end{equation}
where by fixing $X=\frac{1}{2}$ one has that $\Phi_\ell$ is a real scalar in model b).
Regarding the flavor structure of the theory, since the goal is to generate a contribution to $b\to s \mu^+ \mu^-$, we only need couplings to the second and third quark families\footnote{We work in down-quark aligned basis.}, $\lambda^q_3\equiv\lambda_b,\lambda^q_2\equiv\lambda_s$ and to the second generation of leptons $\lambda^\ell_2\equiv\lambda_\mu$. 
The loop-level diagrams responsible for generating the NP contribution to $b\to s\mu^+\mu^-$ are shown in Fig.~\ref{fig:pheno_box} 

\begin{figure}[t!]
\begin{center}
\includegraphics[width=0.48\textwidth]{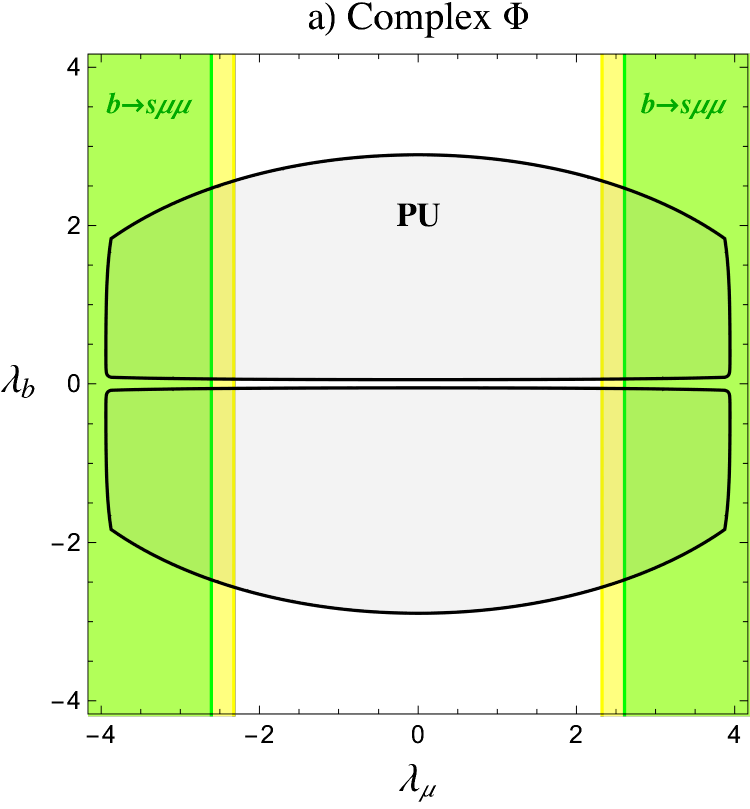}  
\caption{\small In gray we show the regions of the parameter space compatible with PU. In green and yellow we show the regions compatible with the $R_{K^{(*)}}$ anomaly at $1\sigma$ and $2\sigma$ respectively, 
by fitting $\Delta C_9^\mu=-\Delta C_{10}^\mu$ and assuming a common NP mass 
$m_{\rm NP}=1$ TeV. }
\label{fig:LHmodel_a}
\end{center}
\end{figure}

\begin{figure}[t!]
\begin{center}
\includegraphics[width=0.48\textwidth]{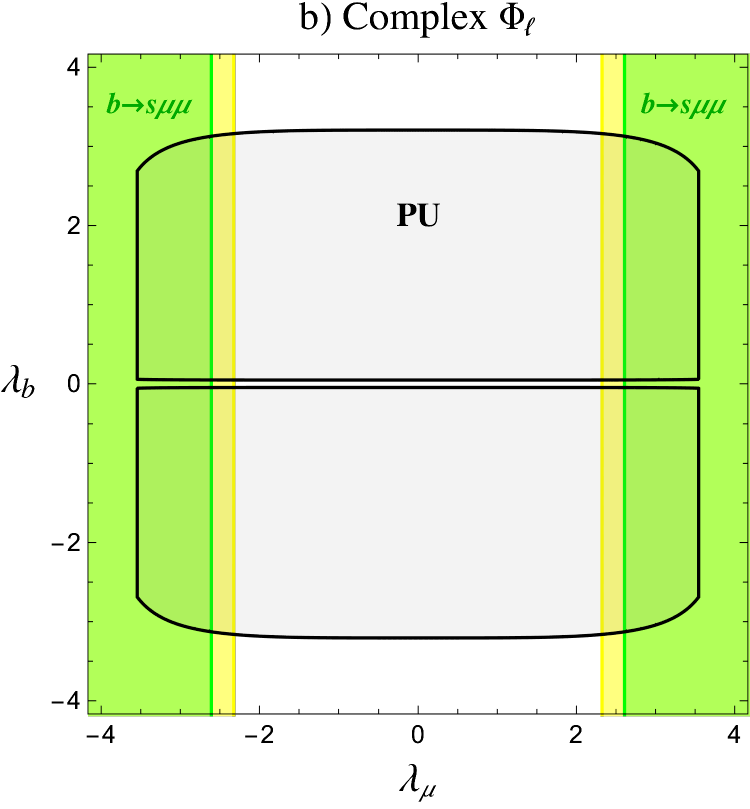}  \hfill
\includegraphics[width=0.48\textwidth]{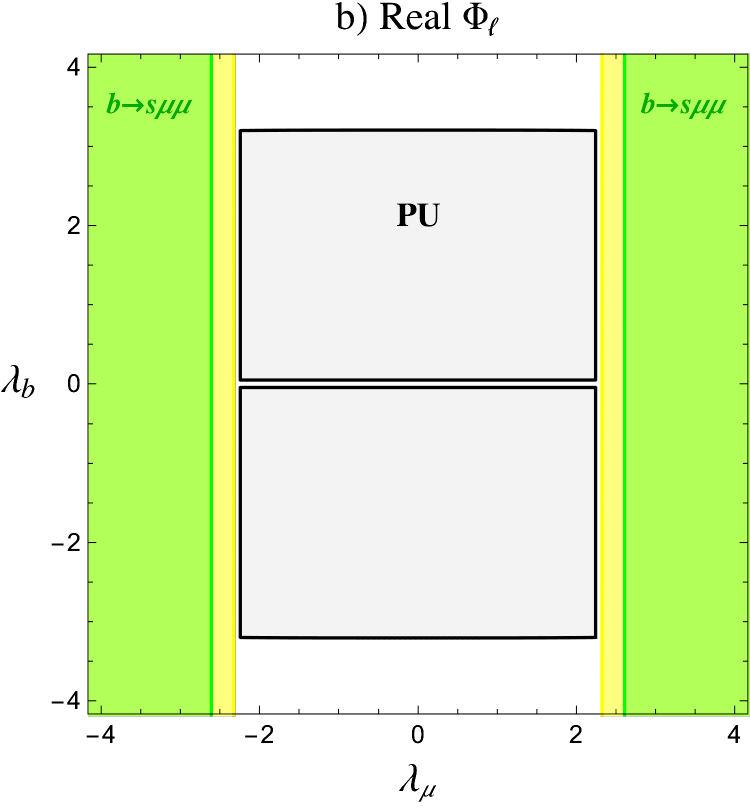} 
\caption{\small In gray we show the regions of the parameter space compatible with PU. In green and yellow we show the regions compatible with the $R_{K^{(*)}}$ anomaly at $1\sigma$ and $2\sigma$ respectively, 
by fitting $\Delta C_9^\mu=-\Delta C_{10}^\mu$ and assuming a common NP mass 
$m_{\rm NP}=1$ TeV.  }
\label{fig:LHmodel_b}
\end{center}
\end{figure}

By fixing for simplicity all the masses of the NP states at a common value $m_{{\rm NP}}\sim {\cal O}({\rm TeV})$,  the most stringent bound for the couplings to quarks comes from $B_s-\bar B_s$ oscillations, where using the result in~\cite{DiLuzio:2019jyq} we get
\begin{equation}
|\lambda_s^*\lambda_b |\le \:0.15\;\frac{m_{\rm NP}}{1\,{\rm TeV}}\,.
\label{eq:SFBsmixingbound}
\end{equation}
This relation can be inserted in the expression for the $\Delta C_9^\mu=-\Delta C_{10}^\mu$ coefficients\footnote{$\Delta C_9^\mu$ is defined as the Wilson coefficient encoding all the NP contributions to the operator $O_9^\mu=\frac{e^2}{16 \pi^2}(\bar{s} \gamma^\nu P_L b)(\bar \mu \gamma_\nu \mu)$, whereas $\Delta C_{10}^\mu$ as the Wilson coefficient encoding all the NP contributions to the operator $O_{10}^\mu=\frac{e^2}{16 \pi^2}(\bar{s}\gamma^\nu P_L b)(\bar \mu \gamma_\nu \gamma_5 \mu)$.  } for reproducing the neutral-current anomaly $R_{K^{(*)}}$ from where one has
\be
\label{eq:SFc9LH}
|\Delta C_9^\mu|=0.34\, |\lambda_s^*\lambda_b| \, |\lambda_\mu|^2 \,\left(\dfrac{1\, {\rm TeV}}{m_{\rm NP}}\right)^2 \, \ ,
\ee 
for both model a) and model b).   By plugging Eq.~\eqref{eq:SFBsmixingbound} into Eq.~\eqref{eq:SFc9LH} one can set a lower bound on the $ |\lambda_\mu|$ coupling~\cite{Arnan:2016cpy} 
\begin{equation}
 |\lambda_\mu|^2 \geq  20\; |\Delta C_9^\mu|\; \dfrac{m_{\rm NP}}{1\, {\rm TeV}} \,,
 \label{eq:c9boxbound}
\end{equation}
where we use the updated 1-dimensional fit $\Delta C_9^{\mu}=-\Delta C_{10}^\mu=-0.41\pm0.07$ in~\cite{Altmannshofer:2021qrr}. Hence, by saturating 
the bound in Eq.~\eqref{eq:SFBsmixingbound}, {\it i.e.}, by imposing $| \lambda_s \lambda_b |\sim \:0.15\;\frac{m_{\rm NP}}{1\,{\rm TeV}}$, we can compute the bound set by PU to see if one can explain at the same time the observed value for $\Delta C_9^\mu$ relevant for the $R_{K^{(*)}}$ anomaly. In order to do so we fix $m_{\rm NP}=1$ TeV, which for $SU(3)_c$ charged NP states is at the edge of exclusion from direct searches at the LHC, and plot the allowed regions from PU in the $(\lambda_\mu,\lambda_b)$ parameter space, accounting for the $\Delta C_9^\mu$ value from~\cite{Altmannshofer:2021qrr}.  We illustrate this for model a) in Fig.~\ref{fig:LHmodel_a} where, as explained before, the scalar is a complex field.
There in green (yellow) we illustrate the regions compatible with the measured value of $\Delta C_9^\mu$ at $1\sigma$ and $2\sigma$ while in gray we show the one compatible with PU.  We see that,  in this case, there is an overlap between the two regions and the $R_{K^{(*)}}$ anomaly can be explained with couplings whose magnitude is compatible with perturbative unitarity.
For the case of model b) we show the results in Fig.~\ref{fig:LHmodel_b} for both the real and complex $\Phi_\ell$ case. In the latter case the results are very similar to the one of ~\ref{fig:LHmodel_a}. On the other side in the case of real $\Phi_\ell$
the PU bounds become more stringent and there is no longer an overlap region where $\Delta C_9^\mu$ can be explained with perturbative couplings. By setting $\lambda_b=0$ for these 3 models, the PU limits for the $\lambda_\mu$ coupling are
\be\label{eq:pheno_model1}
{\rm Model~a)}~~{\rm Complex}~\Phi\quad |\lambda_\mu|<4.0
\qquad
{\rm Model~b)}~~
\begin{cases}
{\rm Complex}~\Phi_\ell\quad |\lambda_\mu|<3.5 \\
{\rm Real}~\Phi_\ell\,~~\qquad |\lambda_\mu|<2.2 \\
 \end{cases}
\ee
where for model a) the bound is obtained from the model in Sec.~\ref{sec:model2}, while for model b) the bounds correspond to the one in Sec.~\ref{sec:model1}.

Since in order to solve the neutral current anomaly we need a coupling to the muons, it is natural to ask if one can reproduce the $(g-2)_\mu$ anomaly and how large the relevant coupling has to be to achieve the correct NP contribution. For the observable $a_\mu\equiv(g-2)_\mu/2$  we consider the recent value of the Fermilab Muon $g-2$ experiment ~\cite{Muong-2:2021ojo} for which one has  a $\sim4.2\sigma$ discrepancy with respect to the SM prediction~\cite{Aoyama:2020ynm}
\be
\Delta a_\mu=a_\mu^{\rm exp}-a_\mu^{\rm SM}=(251\pm59)\times 10^{-11}\,.
\label{eq:gm2}
\ee
We want to see if this anomaly can be explained with the $\lambda_\mu$ coupling in a perturbative regime. To illustrate this we consider the case of model a), since it's the one for which one has a less stringent PU bound. Using the results in~\cite{Arnan:2016cpy}, it turns out that to explain the muon anomaly at  the 1$\sigma$ level by saturating  $\lambda_\mu = 4.0$ as per  Eq.~\eqref{eq:pheno_model1} one needs to have
\begin{equation}
X\left(\dfrac{1\, {\rm TeV}}{m_{\rm NP}}\right)^2 \geq 10.6\, . 
\end{equation}
For a common NP mass of 1\;TeV one needs a quite exotic and large value for the hypercharge
$X=10.6$. In this case one has to assess the validity of the perturbative regime studying scattering processes that involve gauge bosons.
Even more extreme hypercharge values are needed for smaller values of $\lambda_\mu$\footnote{In \cite{Arnan:2016cpy} other representations for the fields are discussed, where for a value of hypercharge $|X|=1$ the muon anomaly can be accounted for with $|\lambda_\mu| \geq 3.7$, although in that case the PU limit also tightens to $|\lambda_\mu| \leq {2.5}$}. 
The need for a large muon coupling in order to explain the $(g-2)_\mu$ anomaly arises because the process needs a chirality flip, which can be obtained in this LH model only via the muon mass term with a contribution proportional to $m_\mu$, which forces the couplings to be too large to account for the anomaly.  This fact could be solved by introducing NP that couples to RH SM muons together with a mixing term among the NP states, since in this case one can generate a chirality flip  proportional to $m_{\rm NP}\gg m_\mu$, that allows for a smaller NP coupling.
 We present some models which include RH couplings in the next Section.

\subsubsection{The inclusion of right-handed couplings}

The possibility of adding RH couplings to scalar-fermion models has been largely  discussed before in the literature, also in the context of DM physics, since for some choices of the field representations one can have suitable DM candidates \cite{Kowalska:2017iqv,Calibbi:2018rzv,Crivellin:2018qmi,Arnan:2019uhr,Arcadi:2021cwg}. 
Introducing a coupling to RH leptons requires at least one new scalar or fermion field. A mixing term among the NP fermion fields can be generated through the interaction with the Higgs boson, while dangerous mixing terms between the Higgs, a SM and a NP field can be forbidden by introducing an extra symmetry like a $\mathcal{Z}_2$ symmetry or a $U(1)$ charge. As explained before, the motivation for introducing RH couplings in these kind of models is to be able to account for the $(g-2)_\mu$ anomaly, while keeping the NP couplings in a perturbative regime. This can be achieved provided that we have a chirality flip contribution bigger to the one proportional to the muon mass.
In a recent work~\cite{Arcadi:2021cwg}, the Authors investigate two models containing a good DM candidate while explaining at the same time the  $b\to s \mu^+ \mu^-$ and the $(g-2)_\mu$ anomalies. In particular one of the two scenarios 
is the extension of model b) of Sec.~\ref{eq:LH} with a real scalar $\Phi_\ell$, whose Lagrangian reads
\begin{equation}
-{\cal L} =  {\lambda^q_i\bar q_L^i  }{\Psi_R}{\Phi_q} + \lambda^\ell_i \bar \ell_L^i  {\Psi_R}{\Phi_\ell} + \lambda^e_i \bar e_R^i {\Psi_L^\prime}{\Phi_\ell} +\lambda^H( \bar \Psi_L{\Psi_R^\prime }{H} + \bar \Psi_R{\Psi_L ^\prime}{H}) + h.c. \ ,
\label{eq:LintbRH}
\end{equation}
where we labeled explicitly the chirality indices $L,R$  in the new fermions $\Psi, \Psi^\prime$. The field quantum numbers that we consider in this case are
\begin{equation}
\begin{array}{ccccc}
\Psi\sim ({\bf 1},{\bf 2},-\frac{1}{2})~~~& \Psi^\prime\sim ({\bf 1},{\bf 1},-1)~~~&\Phi_\ell \sim ({\bf 1},{\bf 1},0)~~~&\Phi_q \sim ({\bf 3},{\bf 1},\frac{2}{3}) \,.
\end{array}
\end{equation}
Again, we restrict our analysis to the case where the flavor structure enforces only couplings to $b$ and $s$ quarks and to muons. Hence we are left with 5 parameters
that will allow us to explain the muon anomalous magnetic moment: $\lambda_b,\lambda_s,\lambda^\ell_\mu,\lambda^e_\mu$ and $\lambda^H$. By fixing, {\emph{e.g.}}, $\lambda^H=0.1$ it turns out that one can explain the $(g-2)_\mu$ anomaly with perturbative couplings. In order to assess whether the neutral-current anomaly can be explained in this scenario while remaining in the perturbative regime
we proceed similarly to the case of the LH scenario and start by saturating the $B_s-\bar B_s$ bound in Eq.~\eqref{eq:SFBsmixingbound}, which fixes the quark coupling combination $|\lambda_b \lambda_s|$. For what concerns the $b\to s\mu^+ \mu^-$ observable, we now have no longer the $\Delta C_9^\mu=-\Delta C_{10}^\mu$  pattern, so that we take the 2D fit result from~\cite{Altmannshofer:2021qrr}
\begin{equation}
\label{eq:C9C10RH}
\Delta C_9^\mu=-0.68\pm0.16 \qquad \Delta C_{10}^\mu=0.24\pm0.13\,, 
\end{equation}
and the $\Delta C_9^\mu$ and $ \Delta C_{10}^\mu$ expressions from~\cite{Arcadi:2021cwg}. 

\begin{figure}[t!]
\begin{center}
\includegraphics[width=0.48\textwidth]{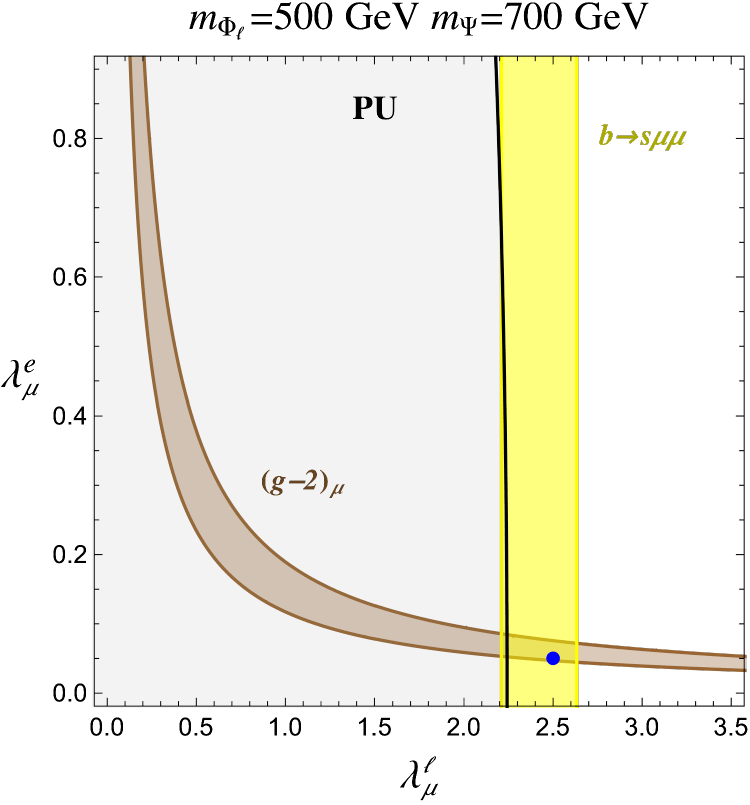}  \hfill
\includegraphics[width=0.48\textwidth]{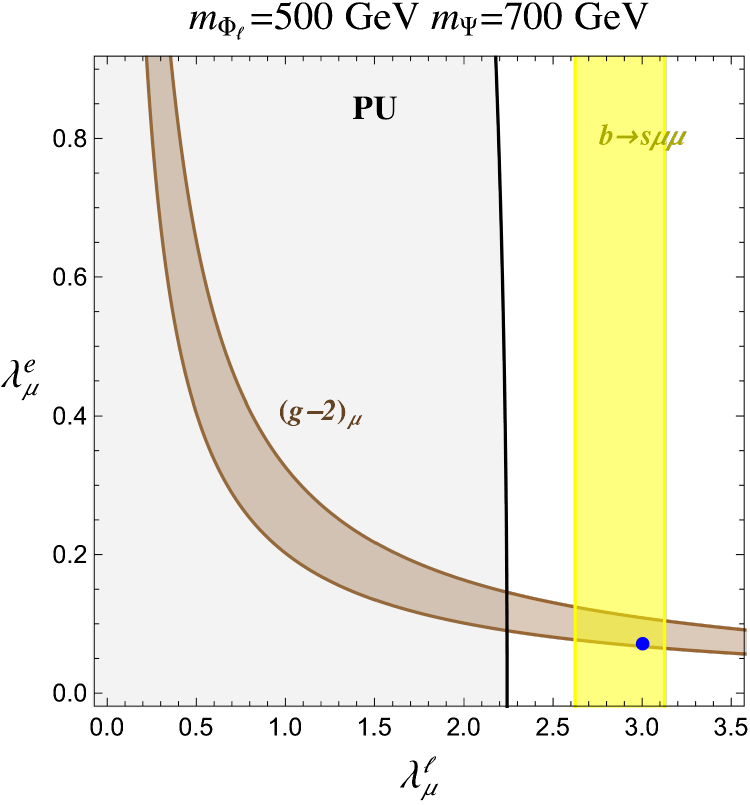} 
\caption{\small 
For benchmark point 1 (left) and benchmark point 2 (right)in gray we show the regions of the parameter space compatible with PU. In brown we show the regions compatible with the $R_{K^{(*)}}$ anomaly at $2\sigma$ while in yellow the region compatible with $(g-2)_\mu$ data at $1\sigma$. For both plot we show in blue the benchmark points derived from ~\cite{Arcadi:2021cwg}. }
\label{fig:SF_RH}
\end{center}
\end{figure}

We then focus on two benchmark points presented in \cite{Arcadi:2021cwg}, that can account for both the $(g-2)_\mu$ and the $R_{K^{(*)}}$ anomalies and the DM relic density, while being compatible with the bounds from direct searches, namely
\begin{itemize}
\item Benchmark 1
\begin{equation}
\lambda_H=0.1,\quad  m_{\Phi_q}=2\text{ TeV},\quad m_{\Psi^\prime}=1\text{ TeV},\quad m_{\Phi_\ell}=0.5\text{ TeV},\quad m_{\Psi}=0.7\text{ TeV}\,,
\end{equation}
\item Benchmark 2
\begin{equation}
\lambda_H=0.1,\quad  m_{\Phi_q}=1.4\text{ TeV},\quad m_{\Psi^\prime}=0.8\text{ TeV},\quad m_{\Phi_\ell}=0.5\text{ TeV},\quad m_{\Psi}=0.7\text{ TeV}\, , 
\end{equation}
\end{itemize}
which uniquely fix the values of  $(\lambda^\ell_\mu,\lambda^e_\mu)$. We then show in gray in Fig.~\ref{fig:SF_RH} the region compatible with the requirement of PU in the $(\lambda^\ell_\mu,\lambda^e_\mu)$ plane, as well as the allowed region for reproducing $(g-2)_\mu$ at 1$\sigma$ and $b\to s\mu\mu$ at 2$\sigma$, which are depicted in brown and yellow respectively. In the figures we also show the $(\lambda^\ell_\mu,\lambda^e_\mu)$ for both benchmark points.
Since this model features a real scalar field and the $\lambda^e_\mu$ needs to be small to satisfy flavor observables, $(g-2)_\mu$ and  $R_{K^{(*)}}$, the PU bound is again dominated  by the results of the model of Sec.~\ref{sec:model1}, which enforces the $|\lambda_\mu| \leq {2.25}$ bound.
From the figures we see that for the first benchmark point there is a tiny region where the $(g-2)_\mu$ anomaly and the $R_{K^{(*)}}$ anomaly can be simultaneously satisfied while being compatible with PU, while for the second benchmark point there is no overlap between the predictions for the various observables while remaining in a perturbative regime.

%%%%%%%%%%%%%%%%
%%%%%%%%%%%%%%%%
\subsection{Scalar leptoquarks}

Scalar LQs are a natural candidate to explain the charged- and neutral-current $R_{D^{(*)}}$ and $R_{K^{(*)}}$ anomalies since they couple quarks to leptons, and are thus an ideal scenario to be tested with the tool of perturbative unitarity.
Among all the scalar LQs the $SU(2)_L$ triplet $S_3$ and singlet $S_1$ are the most robust candidates to explain the anomalies~\cite{Crivellin:2017zlb,Buttazzo:2017ixm,Marzocca:2018wcf,Arnan:2019olv,Crivellin:2019dwb,Saad:2020ihm,Crivellin:2020ukd,Gherardi:2020qhc,DaRold:2020bib,Bordone:2020lnb,Marzocca:2021azj,Marzocca:2021miv}. Under the SM gauge group they transform respectively as $S_3 \sim ({\bf \bar 3}, {\bf 3}, \frac{1}{3})$ and $S_1 \sim ({\bf \bar 3}, {\bf 1}, \frac{1}{3})$. When both LQs are combined so as to explain both the $B-$meson anomalies and the $(g-2)_\mu$, and their mass is set to ${\cal O}(1)\;$TeV, the SM discrepancies can be explained without suffering from PU constraints. However, for higher masses, the couplings are required to be tuned to higher values and perturbativity might be lost.
In a recent work~\cite{Marzocca:2021azj} the Authors have considered the following SM extension
\be\label{eq:lag_Marzocca}
	- {\cal L} = \frac{1}{2}\lambda^\ell_{\alpha \beta} \bar{\ell}^{c,\alpha}_L \varepsilon {\ell}^\beta_L \phi^+ +
	{\bf \lambda}^u_{i\alpha} \bar u_R^{c,i}  e_R^\alpha S_1 + {\bf \lambda}^q_{i\alpha} \bar q_L^{c,i}  \varepsilon \ell_L^\alpha S_1
	 + h.c.~,
\ee
where $\phi^+$ is an $SU(3)_c$ and $SU(2)_L$ singlet scalar with $Y=1$. This model aims at explaining the $B-$anomalies, the anomalous magnetic moment of the muon and the so called Cabibbo Angle Anomaly~\cite{Belfatto:2019swo,Grossman:2019bzp,Crivellin:2020klg} with the following flavor structure:
\be
	{\bf \lambda}^q = \left( \begin{array}{ccc}
					0 & 0 & 0 \\
					0 & 0 &  \lambda_{s\tau}^q \\
					0 & \lambda_{b\mu}^q &  \lambda_{b\tau}^q
					\end{array}\right)~, \qquad
	{\bf \lambda}^u = \left( \begin{array}{ccc}
					0 & 0 & 0 \\
					0 & \lambda_{c\mu}^u  & \lambda_{c\tau}^u \\
					0 & 0 & \lambda_{t\tau}^u 
					\end{array}\right)~, \qquad
	{\bf \lambda} =\left( \begin{array}{ccc}
					0 & \lambda_{e\mu} & 0 \\
					-\lambda_{e\mu} & 0 & \lambda_{\mu\tau} \\
					0 & -\lambda_{\mu\tau} & 0
					\end{array}\right)~ \qquad
	\label{eq:flav_structure_Marzocca} \ .
\ee
Setting $m_{\rm NP}\equiv m_{S_1}=m_{\phi^+}=5.5$ TeV, the best fit point of this model appears in Eq.~(12) of~\cite{Marzocca:2021azj}. Taking the best fit values, each coupling turns out to be in the perturbative regime when considering one of them at the time. However, when considering  the contribution from all the couplings simultaneously, perturbative unitarity is lost.  This is mainly due to the fact that $\lambda_{c\tau}^u$ has to be very large in order to account for the $R_{D^{(*)}}$ anomaly. Explicitly one has~\cite{Azatov:2018kzb,Amhis:2019ckw}
\begin{align}
	\Delta R_D &\approx -0.235\, \lambda_{c\tau}^{u} \lambda^q_{b\tau}\left(\dfrac{1\, {\rm TeV}}{m_{\rm NP}}\right)^2= 0.041\pm 0.029~, \nn \\
	\Delta R_{D^*} &\approx  -0.088\, \lambda_{c\tau}^{u} \lambda^q_{b\tau}\left(\dfrac{1\, {\rm TeV}}{m_{\rm NP}}\right)^2 = 0.037\pm 0.013 \,,
\end{align}
 where $\Delta R_{D^{(*)}}$=$R^{\rm exp}_{D^{(*)}}-R^{\rm SM}_{D^{(*)}}$. 
 
 \begin{figure}[t!]
\begin{center}
\includegraphics[width=0.48\textwidth]{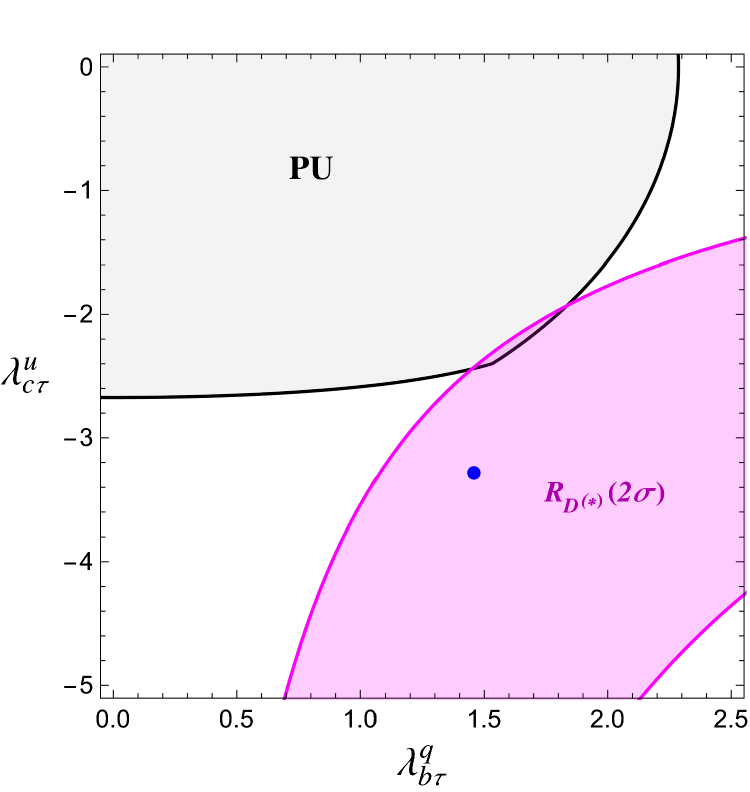}   \hfill
\includegraphics[width=0.47\textwidth]{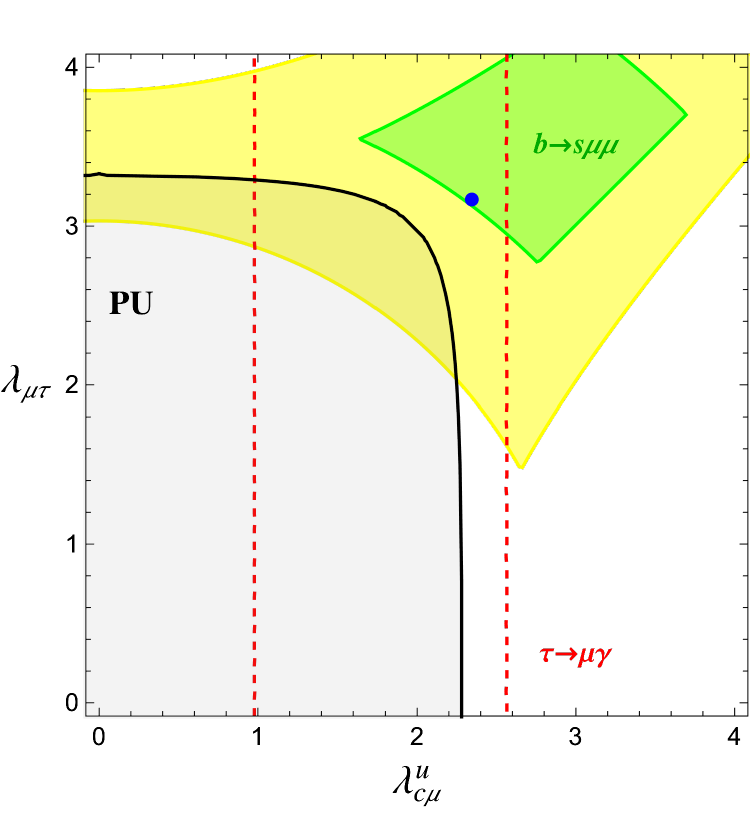} 
\caption{\small {\emph{Left:}} In gray we show the allowed region from PU and in purple the region compatible with $R_{D^{(*)}}$ at 2$\sigma$. The best fit point in~\cite{Marzocca:2021azj} is shown in blue.
{\emph{Right:}}  In gray we show the allowed region from PU while in green and yellow the regions compatible with $R_{K^{(*)}}$ at $1\sigma$ and $2\sigma$ respectively. Between the dashed red lines constraints from $\tau \to \mu \gamma$ are satisfied. The best fit point in~\cite{Marzocca:2021azj} is shown in blue.
}
\label{fig:Marzocca_RD}
\end{center}
\end{figure}

 We then try to see whether it is possible to explain all the anomalies considered in~\cite{Marzocca:2021azj} while remaining in a perturbative regime, without changing the values of $m_{{\rm NP}}=5.5\;$TeV chosen by the Authors.  In the left panel of Fig.~\ref{fig:Marzocca_RD} we show in the $( \lambda_{b\tau}^q , \lambda_{c\tau}^u )$ plane the region compatible with PU, depicted in gray, and the $2\sigma$ region where $R_{D^{(*)}}$ can be reproduced, depicted in purple, where the other couplings are set to their best fit value so that the other relevant anomalies,
 $(g-2)_\mu$ and the Cabibbo Angle,  can be reproduced. We see that there is a small region where $R_{D^{(*)}}$ can be satisfied while being compatible with PU, provided that we lower the value of $|\lambda_{c\tau}^u|$ from the best fit value to $|\lambda_{c\tau}^u|=2.5$ and keep the best fit value for $ \lambda_{b\tau}^q$.
 Thus, by fixing $\lambda_{c\tau}^u=-2.5$  and keeping the other couplings at the best fit as indicated in~\cite{Marzocca:2021azj},  we show in the right panel of  of Fig.~\ref{fig:Marzocca_RD}  in the  $(\lambda_{c\mu}^u,\lambda_{\mu\tau})$ plane the region where the $R_{K^{(*)}}$ anomaly can be reproduced at $1\sigma$ (green) and $1\sigma$ (yellow). There we see that there is compatibility between this requirement and the one of PU, although with a slightly different benchmark point than the one of~\cite{Marzocca:2021azj}. It is important to mention that the coupling 
 $\lambda_{c\mu}^u$ is introduced in the model in order to cancel undesired effects in $\tau \to \mu \gamma$ due to the large value of $\lambda_{c\tau}^u$. For this region
in the right panel of Fig.~\ref{fig:Marzocca_RD} we include the region allowed by $\tau \to \mu \gamma$~\cite{Zyla:2020zbs}, to show the compatibility with this latter measurement.

%%%%%%%%%%%%%%%%
%%%%%%%%%%%%%%%%
\subsection{Yukawa sector in vector leptoquark models}

Other than scalar LQ, a compelling possibility to simultaneously solve the $R_{D^{(*)}}$ and $R_{K^{(*)}}$ anomalies is through a  vector LQ. The most remarkable candidate is the $U_1^\mu$ vector with SM quantum numbers $U_1^\mu\sim({\bf \bar 3}, {\bf 1}, 2/3)$ which has triggered a large theoretical activity aiming at providing an UV completion~\cite{Alonso:2015sja,Calibbi:2015kma,Buttazzo:2017ixm,Kumar:2018kmr,Barbieri:2015yvd,DiLuzio:2017vat,DiLuzio:2018zxy,Bordone:2017bld,Barbieri:2017tuq,Calibbi:2017qbu,Azatov:2018kzb,Blanke:2018sro,Fuentes-Martin:2020bnh,Greljo:2021npi}.
Generally, in order to address the flavor anomalies, the models including $U_1^\mu$ also require the presence of new vector-like fermions and new scalars that couple to the SM via Yukawa couplings which can be constrained by PU considerations.
Here we focus as an example on the model presented in~\cite{DiLuzio:2018zxy}, usually dubbed in the literature as {\emph{4321 model}}, since it possesses a 
gauge symmetry $\mathcal{G}=SU(4)\times SU(3)\times SU(2) \times U(1)$. The Yukawa part of the theory can be divided in a SM-like part and a part which includes the NP fields $\mathcal{L}=\mathcal{L}_{\rm SM-like}+\mathcal{L}_{\rm mix}$. Explicitly
\begin{align}
&- \mathcal{L}_{\rm SM-like} =\bar{q}^\prime_L Y_d H d^\prime_R+\bar{q}^\prime_L Y_u \tilde{H} u^\prime_R+\bar{\ell}^\prime_L Y_e H e^\prime_R +\rm{h.c.}\, \ , \nn \\
&- \mathcal{L}_{\rm mix} =\bar{q}^\prime_L \lambda_q \Omega^T_3 \Psi_R+\bar{\ell}^\prime_L \lambda_\ell \Omega^T_1 \Psi_R+\bar{\Psi}_L( M+\lambda_{15} \Omega_{15} )\Psi_R+\rm{h.c.}\, , 
\end{align}
where we refer to~\cite{DiLuzio:2018zxy} for the field definitions and their quantum numbers under ${\cal G}$. Here we focus on the last term of $\mathcal{L}_{\rm mix}$
which contains the mixing between the new vector-like fermions $\Psi$ and the scalar $\Omega_{15}$, whose quantum numbers under ${\cal G}$ are
\begin{equation}
\begin{array}{ccc}
\Omega_{15}\sim ({\bf 15},{\bf 1},{\bf 1},0)~~~~& \Psi_L \sim ({\bf 4},{\bf 1},{\bf 2},0)~~~~&\Psi_R \sim ({\bf 4},{\bf 1},{\bf 2},0)\ .
\end{array}
\end{equation}
By computing the PU unitarity bound one obtains that the strongest limit is obtained from the $J=1$ channel and reads
\be
\lambda_{15}\lesssim 2.1  \ .
\ee
This is a case where combining different $SU(N)$ factors does not drastically strengthen the bound. Here we have the combination of the model in Sec.~\ref{sec:model1} for $SU(2)$ and the model in Sec.~\ref{sec:model3} for $SU(4)$,  and the possible enhancement in the singlet channel of $J=0$ due to $SU(2)$ structure is cancelled by the $SU(4)$  group factors since the contraction in the $s-$channel of the singlet vanishes. This is the opposite effect of the SM case in Sec.~\ref{sec:sm_gauge_mult} when we considered multiple generations.

Regarding the viability of perturbative couplings of the {\emph{4321 model}}, while in the original work~\cite{DiLuzio:2018zxy} the Authors set $\lambda_{15}\simeq 2.5 $ in order to introduce a mass-spliting between new heavy vector-like quarks and leptons, which would then be in contrast with the perturbative unitarity limit that we have derived,  with the new experimental world averages for the $R_{D^{(*)}}$ and $R_{K^{(*)}}$ anomalies, one can easily lower  the Yukawa coupling to, {\emph{e.g.}}, $\lambda_{15} \sim 2$, while remaining compatible with $\Delta F=2$ observables.
Thus the model is still viable, although the parameters are stretched to the edge of perturbativity according to our criteria.

%%%%%%%%%%%%%%%%%%%
\subsection{Right-handed neutrinos for $B$ anomalies}

\begin{figure}[t!]
\begin{center}
\includegraphics[width=0.48\textwidth]{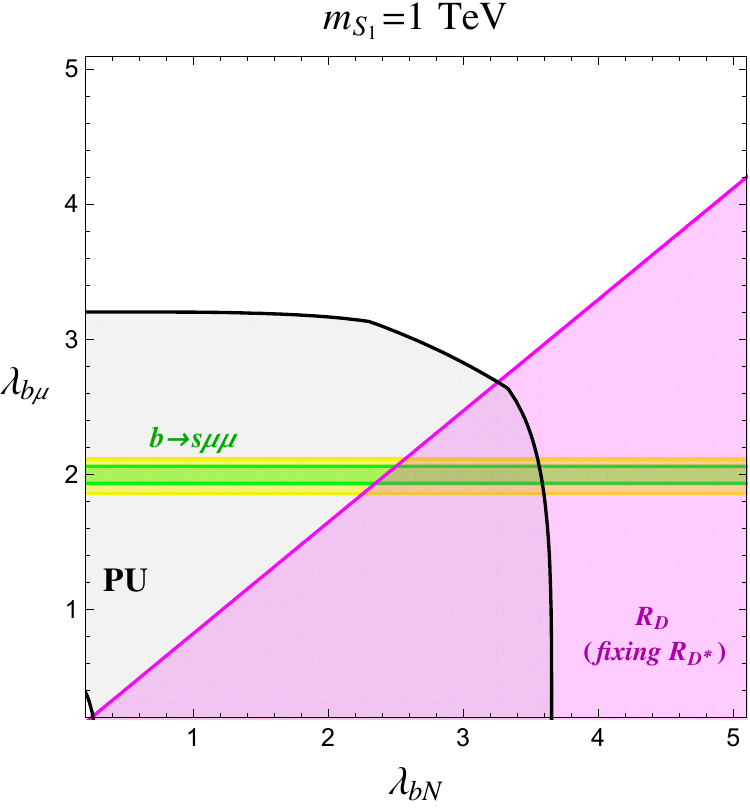}  \hfill
\includegraphics[width=0.48\textwidth]{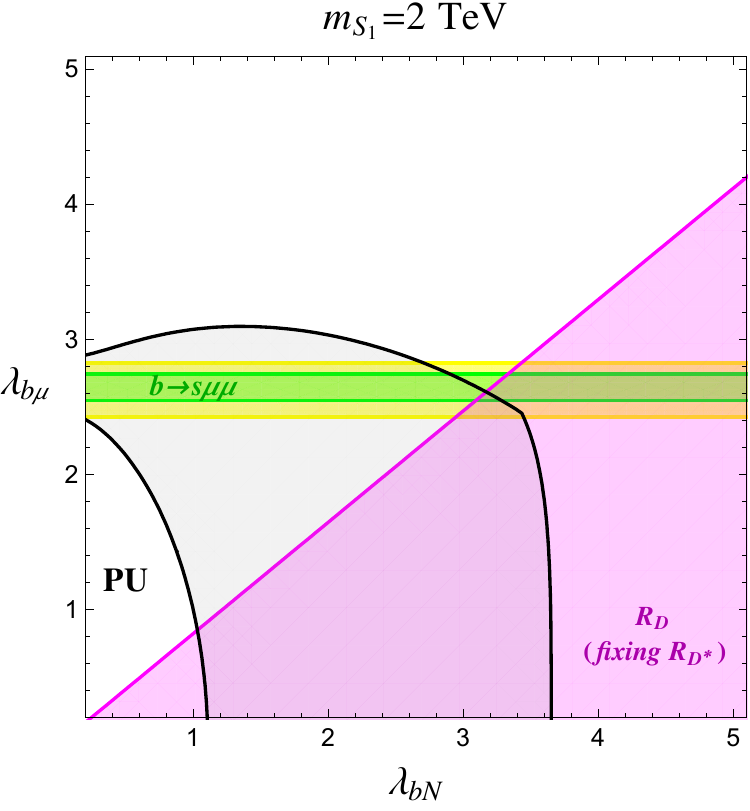} 
\caption{\small In black we show the region compatible with PU of the Yukawa couplings $\lambda^d_{bN}$ and $\lambda^q_{b\mu}$ for a LQ mass of $1\;$TeV (left) and 2\;TeV (right). The brazilian band represents the region compatible with the $R_{K^{(*)}}$  anomaly at 1$\sigma$ and 2$\sigma$, while in the purple region  the measured value $R_{D^{}}$ is reproduced, having fixed $\lambda^u_{c\tau}$ so as to reproduce $R_{D^{(*)}}$. A LQ mass of 1\;TeV is excluded by experimental bounds on $B_s$ mixing.}
\label{fig:RHnu}
\end{center}
\end{figure}

There are also models that can account for the anomalies with the addition of a RH neutrino, thus connecting the flavor tensions with the one of the neutrino mass generation. In~\cite{Azatov:2018kzb} the authors have proposed a model that can address the $R_{D^{(*)}}$ anomaly by adding a new decay channel $B\to D^{(*)} \tau N_R$ into a right-handed sterile neutrino $N_R$ 
while simultaneously solving the 
$R_{K^{(*)}}$ anomaly at one-loop level, through the exchange of a scalar leptoquark $S_1$. The Lagrangian of the theory is 
\be\label{eq:lag_S1}
	-{\cal L} = S_1 (
	{\bf \lambda}^u_{i\alpha} \bar u_R^{c,i}  e_R^\alpha + 
	{\bf \lambda}^d_i  \bar d_R^{ci}  N_R  +
	{\bf \lambda}^q_{i\alpha} \bar q_L^{c,i}  \varepsilon \ell_L^\alpha
	) + h.c.~,
\ee
with the following flavor structure
\be
	{\bf \lambda}^q = \left( \begin{array}{ccc}
					0 & 0 & 0 \\
					0 & \lambda_{s\mu}^q & 0 \\
					0 & \lambda_{b\mu}^q & 0
					\end{array}\right)~, \qquad
	{\bf \lambda}^u = \left( \begin{array}{ccc}
					0 & 0 & 0 \\
					0 & 0 & \lambda_{c\tau}^u \\
					0 & 0 & 0
					\end{array}\right)~, \qquad
	{\bf \lambda}^d = \left( 0, ~ 0, ~ \lambda_{bN}^d \right)^T
	\label{eq:flav_structure_S} \ .
\ee
For what concerns the charged-current anomaly one has~\cite{Azatov:2018kzb,Amhis:2019ckw}
\begin{align}
	 \frac{R_D}{R_D^{{\rm SM}}} &\approx 1 + 0.14 |\lambda_{c\tau}^{u} \lambda^d_{b N}|^2 \left(\dfrac{1\, {\rm TeV}}{m_{S_1}}\right)^4 + 0.19 |\lambda_{c\tau}^{u} \lambda^q_{b\mu}|^2 \left(\dfrac{1\, {\rm TeV}}{m_{S_1}}\right)^4 = 1.137\pm 0.101~, \nn \\
	  \frac{R_{D^*}}{R_{D^*}^{{\rm SM}}} &\approx  1 + 0.14 |\lambda_{c\tau}^{u} \lambda^d_{b N}|^2 \left(\dfrac{1\, {\rm TeV}}{m_{S_1}}\right)^4 + 0.032 |\lambda_{c\tau}^{u} \lambda^q_{b\mu}|^2  \left(\dfrac{1\, {\rm TeV}}{m_{S_1}}\right)^4 = 1.143 \pm 0.057 \ .
\end{align}
 In order to reproduce the neutral current anomaly one has to tune
\be\label{eq:tuning}
\lambda^q_{s\mu} \sim -\frac{V_{cb}}{V_{cs}} \lambda^q_{b\mu}
\ee
in order to avoid violation of lepton flavor universality in $b\to c\ell \nu$ processes, see again ~\cite{Azatov:2018kzb}, where $V$ is the CKM matrix. With this tuning one has that the neutral-current anomaly is reproduced for
\be
|\lambda^q_{b\mu}|^2 \simeq 0.87 + 3.15 \left(\frac{m_{S_1}}{1\;{\rm TeV}} \right)\left(\frac{\Delta C_9^\mu}{-0.41} \right) \ ,
\ee
where we have normalized the expression to the latest best fit for the 
$\Delta C_9^\mu$ coefficient~\cite{Altmannshofer:2021qrr}.
Barring the mass of the RH neutrino, there are four couplings and one mass in this model. One parameter is eliminated by the tuning of Eq.~\eqref{eq:tuning}, while we can eliminate, {\emph{e.g.}}, the value 
of $\lambda^u_{c\tau}$ by asking to reproduce the $R_{D^*}$ anomaly, which is the one with the smaller experimental error. This leaves two independent couplings, $\lambda^d_{bN}$ and $\lambda^q_{b\mu}$, on which we can check the constraints imposed by PU. We show the results in Fig.~\ref{fig:RHnu} for two representative values of the LQ mass. In those figures the region compatible with the PU of the Yukawa couplings is shown in gray, while the brazilian band plot illustrates the region of parameter space that can explain the  $R_{K^{(*)}}$ anomaly at 1$\sigma$ and 2$\sigma$. Finally in purple we show the $2\sigma$ region compatibility for  $R_{D^{}}$, having fixed $\lambda^u_{c\tau}$ so as to reproduce $R_{D^{(*)}}$. Altogether we see that for a LQ mass of 1\;TeV  (left panel) we can simultaneously explain both anomalies while remaining in the perturbative regime. However for this value of the LQ mass, the solutions to the $R_{K^{(*)}}$ anomaly is excluded by the experimental bounds on $B_s$ mixing, see again~\cite{Azatov:2018kzb}. We can restore the compatibility with this measurement by raising the LQ mass up to 2 TeV (right panel), where however now the $\lambda^d_{bN}$ and $\lambda^q_{b\mu}$ couplings are pushed at the edge of the perturbativity.

%%%%%%%%%%%%%%%%%%
%%% 	CONC		%%
%%%%%%%%%%%%%%%%%%

\section{Conclusions}\label{sec:conc}

Yukawa interactions are ubiquitous in NP theories that try to address the shortcomings of the SM and are largely employed in models that try to solve experimental anomalies reported in the recent years in low energy data, as for the case of the muon $(g-2)_\mu$ and semileptonic decays of $B-$mesons. In this paper we have studied the constraints imposed by 
 PU  on generic Yukawa interactions where the fields involved have arbitrary  quantum numbers under an $\prod_i SU(N_i) \otimes U(1)$ group.  
 
 By considering all $2\to 2$ tree-level scatterings in the high-energy limit we have constructed the general form of the partial-wave matrices 
$a^J_{fi}$ and derived compact expressions for the upper limit on the value of the Yukawa interaction up to which perturbation theory can be trusted. 
This has been achieved by computing all the necessary ingredients for building the partial-wave matrix, namely the Lorentz parts of the scattering amplitudes and the group structure factors entering the amplitude themselves. We have started by considering a set of 
phenomenologically relevant toy models with {\emph{Dirac type}} and {\emph{Majorana type}} interactions, where the various fields are only charged under a {\emph{single}}  $SU(N)$ factor, working for concreteness in the case where all the fields transform in the trivial, fundamental or adjoint representation of $SU(N)$ and allowing them to have arbitrary $U(1)$ charges. We have shown how the $SU(N)$ group structure of the interaction can lead to an 
enhancement of the scattering amplitudes  and thus to a tightening of the partial wave unitarity bound, while on the other hand the presence of the $U(1)$ symmetry enforces a selection rule that makes some amplitudes vanish. Interestingly, we obtained that the stronger bound might arise from a partial wave different from $J=0$.

The results obtained for these toy models can then be used as building blocks for more complicated theories, where the the various fields are charged under {\emph{multiple}} $SU(N_i)$ factors. To highlight the strategy we have provided a guided working example, by focusing on the case of the SM quark Yukawa sector. For this case we have also stressed the role that a non trivial flavor structure has in determining the PU bound.  We have then applied our results to various more complicated NP models which solve the aforementioned anomalies in $(g-2)_\mu$ and/or semileptonic $B-$meson decays by postulating 
the existence of new Yukawa interactions. We have highlighted that,
while the proposed theories 
can generally still provide an explanation to these measurements, their models parameters are stretched close to the limit where perturbation theory cannot be trusted and care must be taken in deriving any conclusion.

Finally, the results presented in this paper and illustrated in Figs~\ref{fig:model1},~\ref{fig:model2},~\ref{fig:model3} and \ref{fig:model4},
 are of practical use, and their applicability lies beyond the simple examples presented in the text. While we have restricted only to a limited number of irreducible $SU(N)$ representations under which the various field can transform, 
the expressions that we have derived furnish the necessary ingredients to study the limits imposed by the requirement of PU in a large set of phenomenologically relevant NP theories that present additional Yukawa interactions.

\section*{Acknowledgements}
We thank Luca Di Luzio for useful discussions. The work of MN and PA was supported in part by MIUR under contract PRIN 2017L5W2PT, and by the INFN grant ‘SESAMO’. LA acknowledges support from the Swiss National Science Foundation (SNF) under contract 200021-175940.

%%%%%%%%%%%%%%%%%%
%%% 	APPENDIX		%%
%%%%%%%%%%%%%%%%%%

\appendix
\section{Notation and conventions}\label{app:conventions}

\subsection{Wigner $d-$functions}\label{app:wig}

The small Wigner $d-$functions are defined in the angular momentum basis as
\be
d^j_{m m^\prime}(\theta) = \langle j m^\prime | e^{-i \theta \hat J_y} | j m \rangle  \ ,
\ee
where $\hat J_y$ is the generator of the rotations around the $y-$axis. The explicit expression of these functions used throughout our analysis are
\begin{align}
& d^0_{00}=1 \nn \\
& d^{\frac{1}{2}}_{\frac{1}{2}\frac{1}{2}}=\cos\frac{\theta}{2} \qquad d^{\frac{1}{2}}_{\frac{1}{2}-\frac{1}{2}}=-\sin\frac{\theta}{2}\nn \\
& d^1_{11}=\cos^2\frac{\theta}{2} \qquad d^1_{10}=\frac{\sin\theta}{\sqrt 2} \ ,
\end{align}
with the properties
\be
d^j_{m^\prime m} = (-1)^{m-m^\prime} d^j_{m m^\prime} = d^j_{-m -m^\prime} \ .
\ee

\subsection{Helicity spinor formalism}\label{app:hel}

The fields entering Eq.~\eqref{eq:gen_lag} can explicitly be expanded  in terms of creation and annihilation operators as
\begin{align}\label{eq:fields}
\phi(x) & = \int \dfrac{d^3k}{(2 \pi)^3 \sqrt{2E}} \left[a(k) e^{-ikx}+a(k)^\dag e^{ikx}\right] \ ,  \nn  \\
 \psi_L(x)&=\int \dfrac{d^3k}{(2 \pi)^3 \sqrt{2E}} \left[b_-(k) u_-(k) e^{-ikx}+d^{\dag}_+(k) v_+(k)e^{ikx}\right]  \ , \nn \\
 \psi_L^c(x)&=\int \dfrac{d^3k}{(2 \pi)^3 \sqrt{2E}} \left[b^\dag_-(k) v_-(k)e^{ikx}+d_+(k) u_+(k)e^{-ikx}\right]  \ .
\end{align}
where we choose the spinor basis to be
 \be
u_r(p) = 
\begin{pmatrix}
\sqrt{p \cdot \sigma}\xi_r \\
\sqrt{p \cdot \bar \sigma}\xi_r \\
\end{pmatrix} \ ,
\qquad 
v_s(p) = 
\begin{pmatrix}
\sqrt{p \cdot \sigma}\eta_s \\
- \sqrt{p \cdot \bar \sigma}\eta_s \\
\end{pmatrix} \ ,
\label{eq:spinor_basis}
\ee
where $\sigma^\mu = (\mathbb{1}_2,\sigma^i)$,  $\bar \sigma^\mu = (\mathbb{1}_2,- \sigma^i)$ and $\sigma^i$ are the Pauli matrices. We can choose $\xi$ to be an eigenstate of $\sigma_3$, {\emph{i.e.}} $\xi^+ = (1,0)$ and $\xi^- = (0,1)$ corresponding to spin up and down along the $z$-direction and we fix $\eta^+=(0,1)$ and $\eta^-=(-1,0)$ with the same convention. By building the helicity operator 
\be
\hat \lambda_p = \hat p \cdot S = \frac{\hat p}{2}
\begin{pmatrix}
\sigma^i & 0 \\
0 & \sigma^i \\
\end{pmatrix} \ ,
\ee
one has
\be
\hat \lambda_p u^r(p) = r u^s(p) \ , \qquad \hat \lambda_p v^s(p) = - s v^s(p) \ , 
\ee
where $s,r=\pm 1$ indicate helicity $\pm\frac{1}{2}$ for both particle and antiparticle and where for the antiparticle the helicity is defined with the opposite sign according to standard definitions of helicity spinors, see {\emph{e.g}}~\cite{Chanowitz:1978mv}. 
Then the field $\psi_L$ of Eq.~\eqref{eq:fields} annihilates negative helicity states $\psi_-$ and creates positive helicity states $\psi_+$ while the conjugate field annihilates positive helicity states $\psi_+$ and creates negative helicity states $\psi_-$. To compute the relevant amplitudes we also need rotated spinors that can be built as
\be
u_r(p^\prime) = 
\begin{pmatrix}
R_\theta & 0 \\
0 & R_\theta
\end{pmatrix}
u_r(p) = 
\begin{pmatrix}
\underbrace{R_\theta \sqrt{p\cdot \sigma}R^{-1}_\theta}_{\sqrt{p^\prime\cdot \sigma}} & 0 \\
0 & - \underbrace{R_\theta \sqrt{p\cdot \bar \sigma}R^{-1}_\theta}_{p^\prime\cdot \bar \sigma}
\end{pmatrix}
\begin{pmatrix}
R_\theta \xi_r \\
R_\theta \xi_r
\end{pmatrix} \ ,
\ee
where
\be
R_\theta  =
\begin{pmatrix}
\cos\frac{\theta}{2}\, & - \sin\frac{\theta}{2} \\
\sin\frac{\theta}{2}\, & \cos\frac{\theta}{2}
\end{pmatrix} \ .
\ee
is the rotation matrix in the $x-z$ plane by an angle $\theta$ with respect to the $y-$axes. An analogous expression holds for $v^s(p^\prime)$.

\subsection{Scattering amplitudes in the real scalar basis}

By making the choice of basis where all the fields in Eq.~\eqref{eq:gen_lag} are expressed in terms of their real components, the Lorenz parts of the scattering amplitudes read
\begin{align}\label{eq:amp}
&{\cal{T}}^{s,++++}_{ijkl}  = ({\cal{T}}^{s,----}_{ijkl})^*= -y_{\alpha i j} y_{\alpha k l }^*\ ,  \nn \\ 
&{\cal{T}}^{stu,--++}_{ijkl}  = ({\cal{T}}^{stu,++--}_{ijkl})^* =y_{\alpha i k} y_{\alpha j l }+y_{\alpha i l} y_{\alpha j k }+y_{\alpha i j} y_{\alpha kl } \ , \nn \\
&{\cal{T}}^{su,+0+0}_{i \alpha j \beta}  = ({\cal{T}}^{su,-0-0}_{i \alpha j \beta})^*= -y_{\alpha i k} y_{\beta j k }^* \cos\frac{\theta}{2} - y_{\beta i k} y_{\alpha j k }^*\frac{1}{\cos\frac{\theta}{2}} \ , \nn \\
&{\cal{T}}^{u,+-+-}_{ijkl}  = -y_{\alpha i l} y_{\alpha j k}^*\ , \nn \\
& {\cal{T}}^{tu,00+-}_{ij \alpha \beta}= -({\cal{T}}^{tu,+-00})^*=({\cal{T}}^{tu,00-+})^*= -{\cal{T}}^{tu,-+00}=   y_{\alpha i k} y_{\beta j k}^*\frac{1}{\tan{\frac{\theta}{2}}}-y_{\beta i k} y_{\alpha j k}^*\tan{\frac{\theta}{2}} \ ,
\end{align}
where the $s$, $t$ and $u$ supscripts indicate the Mandelstam channel through which the relative amplitude proceeds.

\section{Other Dirac type theories}\label{app:dirac}

Here we present the results for model 3,4 and 5 in the {\emph{Dirac type}} class.

\subsection{Third model: $\chi\sim {\tiny{\fbox{$\protect\phantom s$} }}_q$, $\eta\sim {\tiny{\fbox{$\protect\phantom s$} }}_{q^\prime}$ $S\sim$Adj$_{q-q^\prime}$ }\label{sec:model3}

In this model $S$ transforms under the adjoint $SU(N)$ representation and is thus a real field if $q=q^\prime$, complex otherwise. We choose as basis
\be
\psi_L = (\chi_{a},\eta^{c,a})^T\ , \qquad
\phi=
\begin{cases}
S^A \qquad\qquad\quad \, \, \,  \, {\rm Real~scalar}\\
(S^A,S^{A*})^T \qquad  {\rm \, Complex~scalar}\\
\end{cases} \ , 
\ee
where the index $a$ and $A$ run from 1 to $N$ and $N^2-1$ respectively. With this choice the Yukawa matrix in the real scalar case reads
\be
{\cal Y}_{\alpha  i j}= y
\begin{pmatrix}
 0_N  &(T^\alpha)_i^{\cdot \, j-N}\\
(T^\alpha)^{j-N}_{~~\cdot~~ i} & 0_N
\end{pmatrix} \ ,
\ee
with $i,j=1,\dots,2N$ and $\alpha=1,\dots,N^2-1$ while in the complex scalar case it is instead
\be
{\cal Y}_{\alpha }= y
\begin{cases}
\begin{pmatrix}
 0_N  &(T^\alpha)_i^{\cdot \, j-N}\\
(T^\alpha)^{j-N}_{~~\cdot~~ i} & 0_N
\end{pmatrix}  \ , \qquad \alpha \le N^2-1 \\
\\
i\begin{pmatrix}
 0_N  &(T^{\alpha-N^2-1})_i^{\cdot \, j-N}\\
(T^{\alpha-N^2-1})^{j-N}_{~~\cdot~~ i} & 0_N
\end{pmatrix} \ , \qquad \alpha > N^2-1
\end{cases} \ ,
\ee
where now  $\alpha=1,\dots,2(N^2-1)$.  In the case of $J=0$ the relevant two particle states decompose again as in Eq.~\eqref{eq:model1_J0_dec}. The relevant group factors for the non zero amplitudes in the real scalar case are
\begin{align}\label{eq:group_model3_J0}
++++
\begin{cases}
{\cal F}^{s,{\rm Adj}}_{\bar \chi \eta \bar \chi \eta} = \frac{1}{2}
\end{cases}
\qquad \qquad
++--
\begin{cases}
{\cal F}^{t, \bf{1}}_{\bar \chi \eta  \chi \bar \eta} = \frac{N^2-1}{2N} \\
{\cal F}^{s, {\rm Adj}}_{\bar \chi \eta  \chi \bar \eta} = \frac{1}{2} \\
{\cal F}^{t, {\rm Adj}}_{\bar \chi \eta  \chi \bar \eta} = -\frac{1}{2N} \\
{\cal F}^{t,{\rm {\bf S}}}_{\eta\eta\chi\chi} = {\cal F}^{u,{\rm {\bf S}}}_{\eta\eta\chi\chi}  =\frac{N-1}{4N} \\
\end{cases} \ .
\end{align}
The partial wave matrix for $J=0$ in the $(\bar\chi \eta, \chi\bar \eta)$ basis\footnote{We use here a compact notation to indicate the basis, where however when considering scattering in representations with dimension greater than one the corresponding two particle states are vectors in that group space.} after integration on the angular variable is
\be\label{eq:model3_j0}
a^{J=0}=\frac{y^2}{16\pi}
\begin{cases}
\begin{pmatrix}
0 & \frac{N^2-1}{2N} \\
\frac{N^2-1}{2N} & 0
\end{pmatrix}\quad\quad\quad\quad\quad  {\rm Singlet} \\
\begin{pmatrix}
-\frac{1}{2} & \frac{N-1}{2N} \\
 \frac{N-1}{2N} & -\frac{1}{2}
\end{pmatrix}\times {\mathbb 1}_{N^2-1}\quad\quad {\rm Adjoint} \\
\begin{pmatrix}
0& \frac{N-1}{2N} \\
 \frac{N-1}{2N} & 0
\end{pmatrix} \times {\mathbb 1}_{\frac{N(N+1)}{2}}
\quad\,\, {\rm Symmetric}
\end{cases}
\ ,
\ee
where $\times$ denotes the Kronecker product.
For $N>1$ the largest eigenvalue of this matrix comes from the singlet channel and is equal to $\frac{y^2}{32\pi}\frac{N^2-1}{N}$, which thus gives the bound
\be
y^2 < 16\pi \frac{N}{N^2-1} \ .
\ee
If $S$ is a complex scalar again the amplitudes in the $\pm\pm\mp\mp$ channels are zero. The only non vanishing scatterings when the matrices in Eq.~\eqref{eq:model3_j0} are diagonal are the ones in the adjoint channel. The largest eigenvalue is $\frac{y^2}{32\pi}$ and the perturbative bound becomes
\be
y^2 <16 \pi \ .
\ee
Moving now to the scattering in the $J=1/2$ partial wave the two particle states now decompose as
\be
+0
\begin{cases}
 \bar\chi S  \sim {\overline{{\tiny{\yng(1)}}}} + {\overline{\bf r_1}} + {\overline{\bf r_2}} \nn \\
 \eta S \sim {\tiny{\yng(1)}} + {\bf r_1} + {\bf r_2}
\end{cases}  \ ,
\ee
where ${\bf r_1}$ and ${\bf r_2}$ are the two irreducible representations arising from the tensor decomposition ${\tiny{\yng(1)}} \times {\rm Adj} = {\tiny{\yng(1)}} + {\bf r_1}+{\bf r_2}$. In tensor component this reads
\begin{align}
& A^i B_k^{\cdot j}  =\nn \\
& = \frac{1}{N^2-1}\left[N A^l B_l^{\cdot j} \delta_k^i - A^l B_l^{\cdot i}\delta_k^j  \right] + \nn \\
& + \frac{1}{2}\left[ A^i B_k^{\cdot j}- A^j B_k^{\cdot i}-\frac{1}{N-1}A^l B_l^{\cdot j}\delta_k^i +\frac{1}{N-1}A^l B_l^{\cdot i} \delta_k^j\right] \nn \\
& + \frac{1}{2}\left[ A^i B_k^{\cdot j}+ A^j B_k^{\cdot i}-\frac{1}{N+1}A^l B_l^{\cdot j}\delta_k^i -\frac{1}{N+1}A^l B_l^{\cdot i} \delta_k^j\right]  \ ,
\label{eq:model3_tensor_deco}
\end{align}
having indicated with $A^i$ and $B_k^{\cdot j}$ two tensors transforming in the antifundamental and adjoint $SU(N)$ representation. The first line of Eq.~\eqref{eq:model3_tensor_deco} indicates the fundamental representation, while the second and third are symmetric and antisymmetric tensors in $i,j$ with null traces with respect to $k$.
 As an example, in the case of $SU(3)$ this reads ${\bf 3}\,\otimes\, {\bf 8}= {\bf 3}\,\oplus\,{\overline{\bf 6}} \oplus\, {\bf{15}}$, and in $SU(4)$ it is ${\bf 4}\otimes {\bf 15}= {\bf 4}\oplus{{\bf 20}} \oplus {\bf{36}}$. The two-particle state in the fundamental representation is easily built as
\be\label{eq:r0}
|S\psi \rangle_{{\tiny{\yng(1)}},a} =\sqrt{\frac{2N}{N^2-1}}  (T^A)_a^{\cdot~i} |S^A \psi_i \rangle  \ .
\ee
For the other two irreducible representations ${\bf r_{1,2}}$ one needs to build by hand the basis for the vector space. Let's start with the representation with the higher dimension ${\bf r_2}$. Here one can split the vector space of the last line of Eq.~\eqref{eq:model3_tensor_deco} in three categories. Tensors where $i\ne j \ne k$, which are  trivially traceless, tensors with $i= j$, $i\ne k$, which again are trivially traceless, and tensors which are traceless but where the null trace arise because of the sum of non zero elements\footnote{This works for $N\ge 3$, since for $SU(2)$ it is not possible to have $i\ne j \neq k$. In order to be able to compute the scattering in the representation $\mathbf{r_2}$ also for $N=2$ ($\mathbf{r_2} = \mathbf{4}$), one can construct e.g. the states with $i = j \neq k$ as
\be
	| S \psi \rangle ^{IK}_{\mathbf{r_2}} = \sqrt{2} \delta^{Ii}\delta^{Ij} (T^A)^{\;j}_k \delta^{kK} | \psi_i S^A \rangle \,.
\ee
This is relevant for instance when computing the bounds for the second model of the Majorana type class.}. 
 One can count the dimensionality of these three categories to be $\frac{N(N-1)(N-2)}{2}$, $N(N-1)$ and $N(N-1)$ respectively, whose sum is $\frac{(N+2)N(N-1)}{2}$, matching the dimensionality of ${\bf r_2}$.  For ${\bf r_1}$, instead, one has that only the tensor with $i\ne j$ are non vanishing due to the antisymmetry in those indices. One can build then two categories for the tensor basis with dimensions $\frac{N(N-1)(N-2)}{2}$ and $N(N-2)$, whose sum is $\frac{N}{2}(N^2-N-2)$ which matches the dimension of ${\bf r_1}$. Note that this representation vanishes for the case of $SU(2)$. In order to compute  the group factor entering the scattering amplitude it's enough to explicitly build only one of this states, since all of them will give the same result. For example we construct the state with unit norm belonging to the first category for ${\bf r_2}$ as
\be\label{eq:r2}
| S \psi \rangle ^{IJK}_{\bf {r_2}} =( \delta^{I i}\delta^{J j}+ \delta^{I j} \delta^{J i} ) (T^A)_k^{\cdot j}\delta^{k K}|  \psi_i S^A\rangle \qquad {\rm with} \qquad I \ne J \ , I\ne K \ , J\ne K \ ,
\ee
where $I,J,K$ label the irreducible representation and range from 1 to $d_{{\bf r_2}}$, $i,j,k$ range from 1 to $N$ and $A$ from 1 to $N^2-1$. Analogously one can build the state in ${\bf r_1}$ of the same category as
\be\label{eq:r1}
| S \psi \rangle ^{IJK}_{\bf {r_1}} =( \delta^{I i}\delta^{J j}- \delta^{I j} \delta^{J i} ) (T^A)_k^{\cdot j}\delta^{k K}|  \psi_i S^A\rangle \qquad I \ne J \ , I\ne K \ , J\ne K \ .
\ee
The group factors entering the amplitudes for $J=1/2$ turn out to be
\begin{align}\label{eq:group_model3_J12}
+0+0
\begin{cases}
{\cal F}^{s,{{\overline{\tiny{\yng(1)}}}}}_{\bar \chi S \bar \chi S} = {\cal F}^{s,{\tiny{\yng(1)}}}_{\eta S \eta S} =\frac{N^2-1}{2N}\\ 
{\cal F}^{u,{{\overline{\tiny{\yng(1)}}}}}_{\bar \chi S \bar \chi S} = {\cal F}^{u,{\tiny{\yng(1)}}}_{\eta S \eta S} =-\frac{1}{2N}\\ 
{\cal F}^{u,{\bf r_1}}_{\bar \chi S \bar \chi S} = {\cal F}^{u,{\bf r_1}}_{\eta S \eta S} =-\frac{1}{2}\\ 
{\cal F}^{u,{\bf r_2}}_{\bar \chi S \bar \chi S} = {\cal F}^{u,{\bf r_2}}_{\eta S \eta S} =\frac{1}{2}\\ 
\end{cases} \ .
\end{align}
In the case of a real scalar $S$ and considering the scattering in the fundamental channel and for the $+0+0$ helicity amplitude, which is then $\eta S\to \eta S$, one has explicitly \begin{align}\label{eq:model3_j12}
a^{J=\frac{1}{2}}_{{\tiny{\yng(1)}}}=
\frac{y^2}{32\pi}
 \int_{-1}^{+1}{\rm d}\cos\theta \;
d^\frac{1}{2}_{\frac{1}{2}\frac{1}{2}}(\theta)\Bigg[   {\cal T}_s^{+0+0} \frac{N^2-1}{2N}-   {\cal T}_u^{+0+0} \frac{1}{2N}\Bigg] \, {\mathbb{1}}_N\ ,
\end{align}
whose largest eigenvalue is $\frac{y^2}{64\pi}\frac{N^2-3}{N}$. For what concerns the scattering in the ${\bf r_{1,2}}$ channels, they all have $\pm \frac{y^2}{32\pi}$ eigenvalues. The bound is thus
\be
y^2 < 16\pi\, {\rm Min}[\frac{2N}{N^2-3},1] \ .
\ee
If the scalar is complex one has that in the $+0+0$ helicity channel the $\eta S\to \eta S$ and $\bar \chi S\to \bar \chi S$ scatterings proceed through $s-$channel, while the $\eta S^* \to \eta S^*$ and $\bar \chi S^*\to \bar\chi S^*$ through $u-$channel. Considering the $s-$channel diagrams in the fundamental channel the eigenvalue is $\frac{y^2}{64\pi}\frac{N^2-1}{N}$ while the $u-$channel diagrams in the ${\bf r_{1,2}}$ channel have eigenvalue $\pm \frac{y^2}{32\pi}$.
The bound is thus
\be
y^2 <16 \pi
\begin{cases}
1\, \,\,\, \, \qquad N \le 2 \\
\frac{2N}{N^2-1} \quad N > 2
\end{cases} \ .
\ee

Finally in the $J=1$ channel, while the $+-$ two-particle states decompose again as in Eq.~\eqref{eq:model1_J0_dec}, the same is not true for $00$, since the scalar field now belongs to the adjoint representation. Here one has the decomposition Adj\,$\otimes\,$Adj$={\bf 1}\oplus{\rm Adj}\oplus{\rm Adj}+\dots$ . Given the irreducible representations that can be built out from  two fermions, only the singlet and adjoints channels are relevant. For real scalar fields the two particle states can be built as
\begin{align}\label{eq:model3_J1_dec}
& |\phi\phi \rangle_1 =  \frac{1}{\sqrt{N^2-1}}\delta^{AB}|\phi^A \phi^B\rangle \nn \\
& |\phi\phi \rangle^A_{{\rm Adj}} =\frac{1}{\sqrt 2}  \frac{f^{ABC}}{\sqrt N}|\phi^B \phi^C\rangle \nn \\
& |\phi\phi \rangle^A_{{\rm Adj}} =\frac{1}{\sqrt 2}\sqrt{\frac{N}{(N^2-4)}} d^{ABC}|\phi^B \phi^{C}\rangle \ ,
\end{align}
where $f^{ABC}$ and $d^{ABC}$ are the antisymmetric and symmetric $SU(N)$ structure constant respectively\footnote{Note that $d^{ABC}$ identically vanishes in $SU(2)$.}.
However, both the singlet and the symmetric adjoint state do not contribute in $J=1$~\cite{Jacob:1959at}. The relevant group factors read
\begin{align}\label{eq:group_model3_J1}
+-+-
\begin{cases}
{\cal F}^{u,{\bf 1}}_{\bar \chi \chi \eta \bar \eta} = {\cal F}^{u,{\bf 1}}_{\eta \bar \eta \bar \chi \chi } = \frac{N^2-1}{2N} \\
{\cal F}^{u,{\rm Adj}}_{\bar \chi \chi \eta \bar \eta} = {\cal F}^{u,{\rm Adj}}_{\eta \bar \eta \bar \chi \chi } = - \frac{1}{2N}\\
{\cal F}^{u,{\rm {\bf S}}}_{\eta\chi\eta\chi} = {\cal F}^{u,{\rm {\bf S}}}_{\bar\chi\bar\eta\bar\chi\bar\eta} =  \frac{N-1}{2N}\\
{\cal F}^{u,{\rm {\bf AS}}}_{\eta\chi\eta\chi} = {\cal F}^{u,{\rm {\bf AS}}}_{\bar\chi\bar\eta\bar\chi\bar\eta} =  \frac{N+1}{2N}\\
\end{cases}
\qquad 
00+-
\begin{cases}
 {\cal F}^{t, \rm{Adj}}_{SS\eta\bar\eta } = - {\cal F}^{t, \rm{Adj}}_{SS\bar\chi \chi}  = -i \frac{\sqrt N}{4}  \\
  {\cal F}^{u \rm{Adj}}_{SS\eta\bar\eta } = - {\cal F}^{u, \rm{Adj}}_{SS\bar\chi \chi }  = i \frac{\sqrt N}{4}  \\
\end{cases} \ .
\end{align}
The channels with the highest eigenvalues are the singlet and adjoint ones, for which the partial wave matrix explicitly reads, in the $(\bar\chi \chi,\eta\bar\eta, S S)$ basis,
\begin{align}
a^{J=1}&=\frac{y^2}{32\pi} \int_{-1}^{+1}{\rm d}\cos\theta \; 
\begin{pmatrix}
0 &  -d^{1}_{11}(\theta) & +2\csc\theta d^{1}_{01}(\theta) \\
- d^{1}_{11}(\theta) & 0 & +2\csc\theta d^{1}_{01}(\theta) \\
-2\csc\theta d^{1}_{10}(\theta) & -2\csc\theta d^{1}_{10}(\theta) & 0
\end{pmatrix}\circ \nn \\
& \circ
\begin{cases}
\begin{pmatrix}
 & \frac{N^2-1}{2N}  & \\
\frac{N^2-1}{2N}  &  &\\
& & 
\end{pmatrix} \quad {\rm Singlet} \\
\begin{pmatrix}
 & -\frac{1}{2N} & +i\frac{\sqrt N}{4} \\
-\frac{1}{2N} & & -i\frac{\sqrt N}{4} \\
-i\frac{\sqrt N}{4} & + i\frac{\sqrt N}{4} & 0
\end{pmatrix}\times {\mathbb 1}_{N^2-1} \quad {\rm Adjoint} \\
\end{cases}
\ ,
\end{align}
where for convenience we have written explicitly the expressions of the ${\cal T}$ amplitudes and where
$2\csc\theta=\frac{1}{\tan\frac{\theta}{2}}+\tan\frac{\theta}{2}$ is the angular part arising from the sum of the $t-$ and $u-$ channel contributions which have a different sign because of the antisymmetry of $f^{ABC}$ entering the amplitude computation. We further note that, as expected, the matrix in the adjoint channel is complex but hermitian, yielding thus real eigenvalues. The strongest bound comes from the adjoint channel for $N<5$ and from the singlet channel for $N>5$ and reads
\be
y^2 <
\begin{cases}
64\pi \frac{N}{1+\sqrt{1+16 N^3}} \, \quad N < 5\\
32 \pi \frac{N}{N^2-1} \,\,\,\qquad \quad N \ge 5
\end{cases} \ .
\ee

\begin{figure}[t!]
\begin{center}
\includegraphics[width=0.55\textwidth]{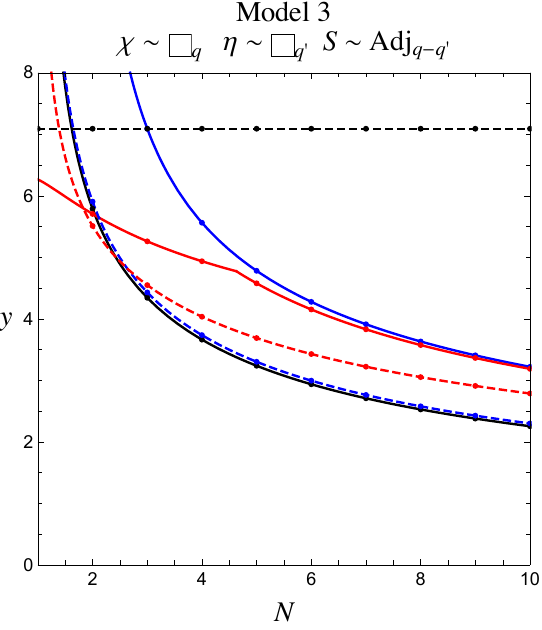} 
\caption{\small PU bounds on the Yukawa couplings $y$ for the model 3 of the {\emph{Dirac type}} class
for $J=0$ (black), $J=1/2$ (blue) and $J=1$ (red).
The solid (dashed) lines correspond to the case of a real (complex) scalar field.  The dashed blue is slightly moved for presentation since it overlays exactly with the solid black.}
\label{fig:model3}
\end{center}
\end{figure}

When the scalar is a complex field one has a non zero contribution also in the singlet channel of the $00^*+-$ scattering, since one can build a non symmetric state, and the scatterings 
proceed now via the $u-$channel. Also in the adjoint channel the contribution from the symmetric state built with the $d^{ABC}$ structure constant, see Eq.~\eqref{eq:model3_J1_dec}, no longer vanishes. The correct normalization for the state is now
\begin{align}
& |S S^* \rangle^A_{{\rm Adj}} = \frac{f^{ABC}}{\sqrt N}|S^B S^{*C}\rangle \ ,  \nn \\
& |S S^* \rangle^A_{{\rm Adj}} =\sqrt{\frac{N}{N^2-4}} d^{ABC}|S^B S^{*C}\rangle \ , 
\end{align}
and the group factors become

\begin{align}\label{eq:group_model3_J1_complex}
00^*+-
\begin{cases}
 {\cal F}^{u, \bf{1}}_{S S^*\eta\bar\eta } =  {\cal F}^{u, \bf{1}}_{S S^*\bar\chi \chi}  =   \frac{1}{2}\sqrt{\frac{N^2-1}{N}} \\
 {\cal F}^{u, \rm{Adj-f}}_{S S^*\eta\bar\eta } = - {\cal F}^{u, \rm{Adj-f}}_{S S^*\bar\chi \chi }  = \frac{i}{2} \sqrt{\frac{N}{2}}  \\
  {\cal F}^{u \rm{Adj-d}}_{S S^*\eta\bar\eta } =  {\cal F}^{u, \rm{Adj-d}}_{S S^*\bar\chi \chi }  =  \frac{1}{2}\sqrt{\frac{N^2-4}{2N}}  \\
\end{cases} \ ,
\end{align}
where we have indicated with Adj$-$f and Adj$-$d the two contributions from the $00$ two-particle states built with the antisymmetric and symmetric $SU(N)$ structure constants respectively. For simplicity we only write the amplitude in the singlet channel, which is the one yielding the stronger limits:
\begin{align}\label{eq:model3_j1complex_sing}
a^{J=1}& =\frac{y^2}{32\pi} \int_{-1}^{+1}{\rm d}\cos\theta \; 
\begin{pmatrix}
0 & d^{1}_{11}(\theta) {\cal T}_u^{+-+-} & -d^{1}_{01}(\theta){\cal T}_u^{00^*+-} \\
d^{1}_{11}(\theta) {\cal T}_u^{+-+-} & 0 & -d^{1}_{01}(\theta){\cal T}_u^{00^*+-}  \\
d^{1}_{10}(\theta){\cal T}_u^{00^*+-} & d^{1}_{10}(\theta){\cal T}_u^{00^*+-}& 0
\end{pmatrix}\circ \nn \\
& \circ
\begin{pmatrix}
 & \frac{N^2-1}{2N}  & \frac{1}{2}\sqrt{\frac{N^2-1}{N}}\\
\frac{N^2-1}{2N}  &  &\frac{1}{2}\sqrt{\frac{N^2-1}{N}}\\
\frac{1}{2}\sqrt{\frac{N^2-1}{N}}&\frac{1}{2}\sqrt{\frac{N^2-1}{N}} & 
\end{pmatrix} \quad {\rm Singlet}
\ .
\end{align}
From the largest eigenvalue of this matrix one obtains the bound
\be
y^2 < 64\pi \frac{N}{N^2-1 +\sqrt{N^4+16 N^3-2N^2-16N+1}} \ .
\ee
Altogether the limits arising from the various partial waves are reported in Fig.~\ref{fig:model3}.

\subsection{Fourth model: $\chi\sim {\tiny{\fbox{$\protect\phantom s$} }}_q$, 
$\eta\sim {\rm Adj}_{q^\prime}$, $S\sim {\tiny{\fbox{$\protect\phantom s$} }}_{q-q^\prime}$  }\label{sec:model4}

In this model $S$ is always a complex scalar field and by fixing the basis
\be
\psi_L = (\chi_{a}, \eta^{c,A})^T \ , \quad \phi = (S_a,S^{*a})^T \ , \quad a=1,\dots,N, \ , \quad A = 1,\dots, N^2-1 \ ,
\ee
where $a$ and $A$ are indices running from 1 to $N$ and $N^2-1$ respectively, the Yukawa matrix reads
\be
{\cal Y}_{\alpha i j } =y
\begin{pmatrix}
0_N & (T^{j-N})_i^{\cdot \,\alpha} \\
(T^{j-N})^{\alpha}_{\cdot \, i} & 0_{N^2-1}
\end{pmatrix} \ .
\ee

As in the previous cases, in the $J=0$ partial wave 
there is no scattering in the $++--$ sector because the scalar is complex, so we focus on the $++++$ channel where only the $\bar \chi \eta \to \bar\chi \eta$ process is non vanishing. We then use the tensor decomposition for ${\tiny{\yng(1)}} \times {\rm Adj} = {\tiny{\yng(1)}} + {\bf r_1}+{\bf r_2}$ of Eq.~\eqref{eq:model3_tensor_deco} and built the two particle states analogously to Eq.~\eqref{eq:r0}, Eq.~\eqref{eq:r2} and Eq.~\eqref{eq:r1}. One obtains non vanishing amplitudes only in the fundamental channels
\begin{align}++++
\begin{cases}
{\cal F}^{s,{\overline{{\tiny{\yng(1)}}}} }_{\bar \chi \eta \bar \chi \eta} =  \frac{N^2-1}{2N} \ ,
\end{cases}
\end{align}
from which one can immediately extract the bound
\be
y^2 < 16\pi\frac{N}{N^2-1} \ .
\ee
In $J=1/2$ we can again focus only on the $+0+0$ helicity amplitude. Here one decomposes the two particle states as
 \be
+0
\begin{cases}
\eta S  \sim  {\tiny{\yng(1)}}+ {\bf r_1} + {\bf r_2}\nn \\
\bar\chi S \sim {\bf 1} + {\rm Adj} \\
\eta S^* \sim \overline{{\tiny{\yng(1)}}} + \overline{{\bf r_1}} + \overline{{\bf r_2}} \\
\bar \chi S^* \sim  {\overline{{\mathbf S}}} + {\overline{{\mathbf{AS}}}}
\end{cases} \ ,
\ee
and the group factors for the non vanishing scatterings are
\begin{align}+0+0
\begin{cases}
{\cal F}^{s,{{\tiny{\yng(1)}}} }_{\eta S\eta S} =  \frac{N^2-1}{2N} \\ 
{\cal F}^{s,{\rm Adj}}_{\bar\chi S\bar\chi S} =  \frac{1}{2} \\
{\cal F}^{u,{{\overline{{\tiny{\yng(1)}}}}} }_{\eta S^*\eta S^*} = - \frac{1}{2N} \\ 
{\cal F}^{u,{\overline{{\bf{r_1}}}} }_{\eta S^*\eta S^*} = -{\cal F}^{u,{\overline{{\bf{r_2}}}} }_{\eta S^*\eta S^*}  = - \frac{1}{2} \\ 
{\cal F}^{u,{\overline{{\bf{S}}}} }_{\bar\chi S^*\bar\chi S^*}   = \frac{N-1}{2N}\\
{\cal F}^{u,{\overline{{\bf{AS}}}} }_{\bar\chi S^*\bar\chi S^*}   = \frac{N+1}{2N}\end{cases} \ .
\end{align}
Also in this case the channel yielding the stronger limits depends on the value of $N$. One obtains
\be
y^2 < 16 \pi 
\begin{cases}
\frac{N}{N+1} \quad N \le 3\\
\frac{2N}{N^2-1} \quad  N \ge 3\\
\end{cases}
\ee
where the first comes from the scattering in the antisymmetric channel while the second from the one in the fundamental.

\begin{figure}[t!]
\begin{center}
\includegraphics[width=0.55\textwidth]{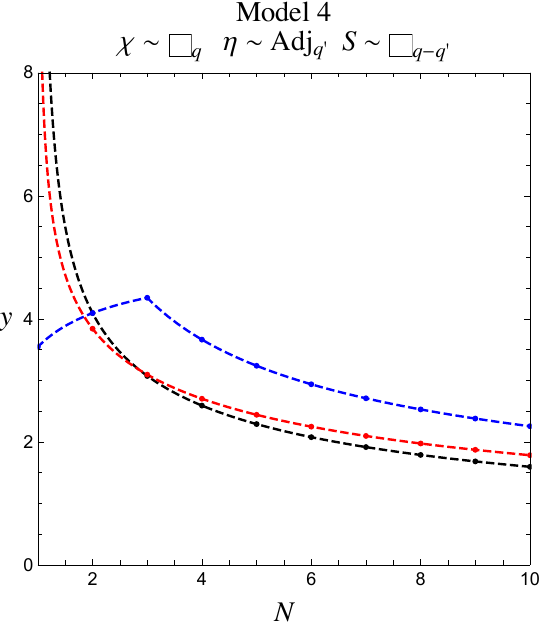} 
\caption{\small PU bounds on the Yukawa couplings $y$ for the model 4 of the {\emph{Dirac type}} class
for $J=0$ (black), $J=1/2$ (blue) and $J=1$ (red).
In this case the scalar is always a complex field. }
\label{fig:model4}
\end{center}
\end{figure}

Finally, for $J=1$ we can decompose the two-particle states as
 \be
+-
\begin{cases}
\bar\chi \chi \sim 1+ {\rm Adj} \nn \\
\bar\eta \eta \sim 1 + {\rm Adj} + {\rm Adj} + \dots \nn \\
\eta \chi   \sim  {\tiny{\yng(1)}} + {\bf r_{1}} + {\bf r_{2}}\nn \\
\bar\chi \bar\eta   \sim  {\overline{{\tiny{\yng(1)}}}}  + {\overline{{\bf r_{1}}}} + {\overline{{\bf r_{2}}}}\\
\end{cases}
\qquad
00
\begin{cases}
S S \sim {\bf S}+ {\bf AS} \nn \\
S S^* \sim 1+{\rm Adj} \nn \\
S^* S^* \sim {\bar{\bf S}}+\overline{{\bf AS}}
\end{cases} \ ,
\ee
and the group factors for the non vanishing amplitudes are

\begin{align}\label{eq:group_model3_J1}
+-+-
\begin{cases}
{\cal F}^{u,{\bf 1}}_{\bar \chi \chi \eta \bar \eta} = {\cal F}^{u,{\bf 1}}_{\eta \bar \eta \bar \chi \chi } =  \frac{1}{2}\sqrt{\frac{N^2-1}{N}}\\
{\cal F}^{u,{\rm Adj-f}}_{\eta\bar\eta \bar\chi\chi}=-{\cal F}^{u,{\rm Adj-f}}_{ \bar\chi\chi \eta\bar\eta}  = -\frac{i}{2}\sqrt{\frac{N}{2}} \\
{\cal F}^{u,{\rm Adj-d}}_{\eta\bar\eta \bar\chi\chi}={\cal F}^{u,{\rm Adj-d}}_{ \bar\chi\chi \eta\bar\eta}  =  \frac{1}{2\sqrt 2}\sqrt{\frac{N^2-4}{N} } \\
{\cal F}^{u,{\tiny{\yng(1)}}}_{\eta\chi\eta\chi} = -\frac{1}{2N}\\
{\cal F}^{u,{\bf r_1}}_{\eta\chi\eta\chi} = - {\cal F}^{u,{\bf r_2}}_{\eta\chi\eta\chi} = -\frac{1}{2} \\
\end{cases}
\qquad \qquad
00^*+-
\begin{cases} 
{\cal F}^{u,{\bf 1}}_{S S^*\bar \chi \chi } = \frac{N^2-1}{2N} \\
{\cal F}^{u,{\bf 1}}_{S S^*\bar \chi \chi } = \frac{1}{2}\sqrt{\frac{N^2-1}{N}} \\
{\cal F}^{u,{\rm Adj}}_{ S S^*\bar \chi \chi} = -\frac{1}{2N}\\
{\cal F}^{u,{\rm Adj-f}}_{S S^*\eta\bar \eta } = \frac{i}{2}\sqrt{\frac{N}{2}} \\
{\cal F}^{u,{\rm Adj-d}}_{S S^*\eta\bar \eta } = \frac{1}{2\sqrt 2}\sqrt{\frac{N^2-4}{N}}\\
\end{cases} \ .
\end{align}
For brevity we report only the partial wave in the singlet channel, which is the one giving the most stringent bound, which reads, in the basis $(\bar \chi \chi, \eta\bar \eta, S S^*)$,
\be\label{eq:model4_j1complex_sing}
a^{J=1}=\frac{y^2}{32\pi} \int_{-1}^{+1}{\rm d}\cos\theta \; 
\begin{pmatrix}
0 & d^{1}_{11}(\theta) {\cal T}_u^{+-+-}& -d^{1}_{01}{\cal T}_u^{00^*+-} \\
d^{1}_{11}(\theta) {\cal T}_u^{+-+-} & 0 &  -d^{1}_{01}{\cal T}_u^{00^*+-} \\
 d^{1}_{10}(\theta) {\cal T}_u^{00^*+-} &  d^{1}_{10} {\cal T}_u^{00^*+-} & 0
\end{pmatrix}
\circ
\begin{pmatrix}
 &  \frac{1}{2}\sqrt{\frac{N^2-1}{N}}  & \frac{N^2-1}{2N}\\
 \frac{1}{2}\sqrt{\frac{N^2-1}{N}}  &  &\frac{1}{2}\sqrt{\frac{N^2-1}{N}}\\
\frac{N^2-1}{2N}&\frac{1}{2}\sqrt{\frac{N^2-1}{N}} & 
\end{pmatrix}  
\ .
\ee
The eigenvalues of this matrix have a complicated form and we report the numerical results in Fig.~\ref{fig:model4}, together with the limits from the other partial waves.

\subsection{Fifth model: $\chi\sim {\tiny{\fbox{$\protect\phantom s$} }}_q$, 
$\eta\sim  {\overline{{\tiny{\fbox{$\protect\phantom s$} }}}}_{q^\prime}$, $S\sim{\overline{ {\tiny{\fbox{$\protect\phantom s$}}} }}_{q-q^\prime}$  }\label{sec:model5}

This last model is only present in the case of $SU(3)$ since it's possible to build a singlet from three fundamentals by using the three-dimensional Levi-Civita tensor\footnote{We fix $\varepsilon_{123}=1$.}.  With the choice of basis
\be
\psi_L = (\chi_{a}, \eta^{c,a})^T \ , \quad \phi = (S_a,S^{a*})^T  \ , 
\ee
where $a$ runs from $1$ to $N$ the Yukawa matrix reads
\be
{\cal Y}_{\alpha i j} = y
\begin{pmatrix}
0_N & \varepsilon_{\alpha,i,j-N} \\
- \varepsilon_{\alpha,i-N,j} & 0_N
\end{pmatrix} \ .
\\
\ee
In the $J=0$ partial wave only the $\pm\pm\pm\pm$ scatterings proceeding through the antisymmetric channel are non zero, due to the antisymmetry of $\varepsilon$. The group factor is simply $2 $ and the bound turns out to be
\be
y^2 < 4\pi \ .
\ee
For $J=1/2$ the two particle states decompose as
 \be
+0
\begin{cases}
\eta S  \sim  {\rm{\bf{\overline{S}}}}+{\rm{\bf \overline{AS}}} \nn \\
\eta S^*  \sim {\bf 1}+{\rm Adj}\nn \\
\bar\chi S \sim {\rm{\bf{\overline{S}}}}+{\rm{\bf \overline{AS}}}  \nn \\
\bar\chi S^*  \sim{\bf 1}+{\rm Adj} \nn \\
\end{cases} \ ,
\ee
and the group factors are
\begin{align}+0+0
\begin{cases}
{\cal F}^{u,{{{\bf 1}}} }_{\eta S^*\eta S^*} = {\cal F}^{u,{{{\bf 1}}} }_{\bar\chi S^*\bar\chi S^*}  =(N-1)\\ 
{\cal F}^{u,{{{\rm Adj}}} }_{\eta S^*\eta S^*} = {\cal F}^{u,{{{\rm Adj}}} }_{\bar\chi S^*\bar\chi S^*}  = -1\\ 
{\cal F}^{s,{\overline{{\bf AS}}} }_{\eta S\eta S} = {\cal F}^{s,{\overline{{\bf AS}}} }_{\bar\chi S\bar\chi S}= 2 \\ 
\end{cases} \ ,
\end{align}
where now $N=3$.
The highest eigenvalues clearly arise from the singlet and antisymmetric channels which lead again to the bound
\be
y^2 < 4\pi \ .
\ee
Finally for $J=1$  all the two-fermion states decompose as $ \sim  {\bf 1}+ {\rm Adj} $ and the same is true for the $S S^*$ scalar state, which is the only one which leads to non zero amplitudes. The group factors for $+-+-$ are all $\pm(N-1)$ for scatterings in the singlet channel and $\pm1$ for scatterings in the adjoint one.  On the other side in the $00+-$ helicity channel for the group factors one obtains $\pm 2$ in the singlet channel and $\pm 1$ for the adjoint one. The most stringent bound is then obtained from the scattering among singlets where one has, in the $( \eta\bar \eta, \eta \chi, \bar\chi\bar\eta, \bar\chi\chi,S S^*)$ basis and after the angular integration,
\begin{align}
a^{J=0}_{{\bf 1}}& =\frac{y^2}{16\pi} 
\begin{pmatrix}
& & &-1 &-\sqrt{2} \\
& 1& & & \\
& & 1& & \\
-1& & & & -\sqrt{2}\\
-\sqrt 2 & & & -\sqrt 2 & \\
\end{pmatrix} \ , 
\end{align}
which gives the bound
\be
y^2 <  \frac{16}{1+\sqrt {17}}\pi \ ,
\ee
which is the most stringent among the various partial waves.

\clearpage\newpage
\bibliographystyle{JHEP}
{\footnotesize
\bibliography{biblio}}

\end{document}